\pdfoutput=1  

 \documentclass[aps,groupedaddress,floatfix,preprintnumbers,nofootinbib,eqsecnum]{revtex4}
\usepackage[T1]{fontenc}
\usepackage[utf8]{inputenc}
\usepackage{amssymb,amsmath,multirow}
\usepackage{graphics,graphicx,epsfig,ulem}
\usepackage{float}
\usepackage{multirow}
\usepackage{longtable}
\usepackage{color}
\usepackage{xcolor}
\usepackage{slashed}
\usepackage[toc,page]{appendix}
\usepackage{hyperref}

\newcommand{\gym}{g_{\scriptscriptstyle Y\hspace{-0.1cm}M}}

\def\xx{\hspace{0.3cm}}
\def\xxl{\hspace{0.29cm}}
\def\xxh{\hspace{0.24cm}}
\def\yy{\hspace{0.11cm}}
\def\yyr{\hspace{-0.02cm}}

\def\beq{\begin{equation}}
\def\eeq{\end{equation}}
\def\beqn{\begin{eqnarray}}
\def\eeqn{\end{eqnarray}}
\newcommand{\newc}{\newcommand}
\newc{\gsim}{\lower.7ex\hbox{$\;\stackrel{\textstyle>}{\sim}\;$}}
\newc{\lsim}{\lower.7ex\hbox{$\;\stackrel{\textstyle<}{\sim}\;$}}

\def\ie{{\it i.e.}\/}

\def\IR{\relax{\rm I\kern-.18em R}}
 \font\cmss=cmss10 \font\cmsss=cmss10 at 7pt
\def\IQ{\relax{\rm I\kern-.18em Q}}
\def\IZ{\relax\ifmmode\mathchoice
 {\hbox{\cmss Z\kern-.4em Z}}{\hbox{\cmss Z\kern-.4em Z}}
 {\lower.9pt\hbox{\cmsss Z\kern-.4em Z}}
 {\lower1.2pt\hbox{\cmsss Z\kern-.4em Z}}\else{\cmss Z\kern-.4em Z}\fi}

\begin{document}

\title{GUT Precursors and Entwined SUSY:\\ The Phenomenology of Stable Non-Supersymmetric Strings\\}
\author{ Steven Abel$^{1}$\footnote{E-mail address:
      {\tt s.a.abel@durham.ac.uk}},
       Keith R. Dienes$^{2,3}$\footnote{E-mail address:
      {\tt dienes@email.arizona.edu}},
      Eirini Mavroudi$^{1}$\footnote{E-mail address:
      {\tt irene.mavroudi@durham.ac.uk}}}
\affiliation{
     $^1\,$IPPP, Durham University, Durham, DH1 3LE, United Kingdom\\
     $^2\,$Department of Physics, University of Arizona, Tucson, AZ  85721  USA \\
     $^3\,$Department of Physics, University of Maryland, College Park, MD  20742  USA }

\begin{abstract}
\noindent
Recent work has established a method of constructing non-super\-sym\-metric string models that are stable, with near-vanishing one-loop dilaton tadpoles and cosmological constants.  This opens up the tantalizing possibility of realizing stable string models whose low-energy limits directly resemble the Standard Model rather than one of its supersymmetric extensions.  In this paper we consider the general structure of such strings and find that they share two important phenomenological properties.  The first is a so-called ``GUT-precursor'' structure in which new GUT-like states appear with masses that can be many orders of magnitude lighter than the scale of gauge coupling unification.   These states allow a parametrically large compactification volume, even in weakly coupled heterotic strings,
and in certain regions of parameter space 
can give rise to dramatic collider signatures which serve as ``smoking guns'' 
for this overall string framework.
The second is a residual ``entwined-SUSY'' (or {\it e}\/-SUSY) structure for the matter multiplets 
in which different multiplet components carry different horizontal $U(1)$ charges.  
As a concrete example and existence proof of these features, we present a heterotic string model that contains the fundamental building blocks of the Standard Model such as the Standard-Model gauge group, complete chiral generations, and Higgs fields --- all without supersymmetry.  Even though massless gravitinos and gauginos are absent from the spectrum, 
we confirm that 
this model has an exponentially suppressed one-loop dilaton tadpole and 
displays both the GUT-precursor and {\it e}\/-SUSY structures.
We also discuss some general phenomenological properties of {\it e}\/-SUSY, such 
as cancellations in radiative corrections to scalar masses,
the possible existence of a corresponding approximate moduli space,
and the prevention of rapid proton decay.
               
\end{abstract}

\maketitle 


\tableofcontents
\section{Introduction}

Most approaches to string phenomenology
have historically proceeded 
under the assumption that the Standard Model (SM) ultimately becomes supersymmetric at a higher energy scale
parametrically near the electroweak symmetry-breaking scale. One then 
attempts to realize the 
resulting supersymmetric theory as the low-energy limit
of a supersymmetric string.
This approach was motivated by many factors.
While bottom-up factors included a strong belief in the existence of weak-scale supersymmetry,
a critical top-down factor was the fact that non-supersymmetric strings are generally
unstable, with large one-loop dilaton tadpoles.
The existence of such tadpoles destabilizes these strings, and thus renders them inconsistent
in a way that does not arise for supersymmetric strings.

In recent work~\cite{Abel:2015oxa}, we advocated a new approach to this problem.
Specifically, even though
non-supersymmetric strings are generally unstable,
they may nevertheless be {\it metastable} --- \ie, endowed with lifetimes that are
large compared with the age of the universe.
Indeed this metastability can be arranged not through the existence
of a potential barrier through which an eventual non-perturbative tunneling might 
occur, but simply by having one-loop dilaton tadpoles whose
values --- although non-zero --- are exponentially suppressed.
Thus, while such strings do not necessarily sit at true minima of the dilaton potential,
the potential slopes that they experience are exponentially suppressed.
Such strings therefore remain effectively stable at their original
locations for all relevant cosmological timescales.

In Ref.~\cite{Abel:2015oxa}, we demonstrated how such metastable strings may be constructed
within the perturbative heterotic framework.
Moreover, as we demonstrated, the low-energy limits of these strings may even resemble the Standard Model or one 
of its grand-unified extensions~\cite{Abel:2015oxa}.
This then opens up the possibility of developing a fully {\it non-supersymmetric}\/ string  phenomenology ---
one in which the Standard Model itself is realized directly as the low-energy limit of a non-supersymmetric string.
Indeed, such models take the general form of a low-energy theory in which supersymmetry (SUSY) is broken at arbitrarily
high scales, yet with a one-loop cosmological constant and 
dilaton tadpole that are exponentially suppressed --- all capped off with a self-consistent ultraviolet (UV) completion which is entirely non-supersymmetric. 
Indeed, as discussed in Ref.~\cite{Abel:2016hgy},
although these theories admit low-energy descriptions in terms
of four-dimensional effective field theories with broken supersymmetry,
they are never even approximately supersymmetric in four dimensions. 

In this paper, we take the next steps in exploring the phenomenological implications of this approach.
In particular, because our construction necessarily involves large-volume compactifications, 
one pressing issue concerns the behavior of the gauge couplings --- especially if we require perturbativity
both at the electroweak scale as well as in the UV limit.        
As we shall discuss, this requires that our strings exhibit a variant of the so-called ``GUT precursor'' structure
originally proposed in Refs.~\cite{Dienes:2002bg,Dienes:2004rt}.~
Tightly coupled with this, we shall also argue that the (chiral) 
matter fields of such strings exhibit a so-called ``entwined SUSY''
or {\it e}\/-SUSY in which these states and their would-be superpartners have different charges under a horizontal $U(1)$ symmetry.  This horizontal $U(1)$ symmetry is thus non-trivially ``entwined'' with  
the same physics that renders the theory non-supersymmetric and also breaks the GUT symmetry.
In this connection, we note that 
entwined SUSY is reminiscent of the so-called ``folded SUSY'' framework~\cite{Burdman:2006tz,Cohen:2015gaa}
in which would-be superpartners have different $SU(3)$ charges. 
Indeed, it might even be possible to incorporate folded SUSY or its variants into our construction.
However, as we shall see, it is actually 
entwined SUSY which unavoidably emerges 
from our overall stable-string construction
and which even serves as one its {\it predictions}\/.

The construction of non-supersymmetric strings has been explored by a number of authors in recent years 
(see, for example, Refs.~\cite{
Angelantonj:2014dia,
Hamada:2015ria,
Nibbelink:2015ena,
Florakis:2015txa,
Ashfaque:2015vta,
Blaszczyk:2015zta,
Nibbelink:2015vha,
Angelantonj:2015nfa,
Satoh:2015nlc,
Athanasopoulos:2016aws,
Sugawara:2016lpa,
Kounnas:2016gmz,
Florakis:2016ani,
Satoh:2016izo,
Kounnas:2017mad,
Florakis:2017ecd,
Faraggi:2017cnh,
Coudarchet:2017pie,
Mourad:2017rrl}).
This growing literature indicates an increasing interest in this subject, presumably motivated not only  
by the apparent experimental absence of supersymmetry at the Large Hadron Collider but also by the intrinsically different theoretical behavior of strings within this hitherto largely unexplored region of the string landscape.
However, within this literature, what distinguishes our work is its focus on the fundamental {\it stability}\/ properties of such strings, at least as far as their dilaton tadpoles are concerned.  Indeed, the presence of a non-zero dilaton tadpole indicates that the fundamental string vacuum is unstable.   It is thus only by concentrating on string models with vanishing or near-vanishing dilaton tadpoles that one can be assured of working in string vacua
whose stability properties resemble those of their supersymmetric cousins.
Of course, just as for supersymmetric strings, there will always remain further moduli which also require stabilization through either string-theoretic or field-theoretic means.
However, we view the dilaton tadpole as uniquely problematic in the construction of non-supersymmetric strings, as the existence of such a tadpole is the direct hallmark of the breaking of supersymmetry.
This problem must therefore be tackled at the outset.
Indeed, it is only after the effective cancellation of this tadpole
that we can proceed to consider the development of a non-supersymmetric 
string phenomenology on a par with that of strings with spacetime supersymmetry.

This paper is organized as follows.
First, in Sect.~\ref{sec:framework}, we review our general framework~\cite{Abel:2015oxa} for the construction of non-supersymmetric heterotic strings with exponentially suppressed one-loop dilaton tadpoles.
Then, in Sect.~\ref{entwined}, we discuss how and why these strings inevitably give rise to not only GUT precursors but also an entwined SUSY  --- observations  that form the central core of this paper.
In Sect.~\ref{section4} we then proceed to construct a self-consistent non-supersymmetric heterotic 
string model which exhibits all of these properties. 
Our aim is to present not only a concrete example and existence proof 
of these features within the context of a fully self-consistent
string model, but also to demonstrate that these features can coexist with other fundamental 
phenomenological 
building blocks of realistic string models
such as the Standard-Model gauge group, complete chiral generations, and Higgs fields --- all in a
stable, non-supersymmetric setting.  
In Sect.~\ref{pheno}, we then briefly discuss several other phenomenological aspects of metastable string models
that result from their GUT-precursor and {\it e}\/-SUSY structures.   These include
cancellations in radiative corrections to scalar masses,
the possible existence of a corresponding approximate moduli space,
and the prevention of rapid proton decay.
Finally, in Sect.~\ref{discussion}, we discuss a variety of open topics and future directions related to our work.
Details pertaining to a calculation in Sect.~\ref{section4} are collected in an Appendix.

\section{Stable non-supersymmetric strings:   ~Basic framework
\label{sec:framework}}

We begin by briefly summarizing the framework described in Ref.~\cite{Abel:2015oxa} for constructing closed, non-supersymmetric string theories with exponentially suppressed one-loop dilaton tadpoles.  
All of the strings we consider in this paper will be members of this class.

There are two critical features which define this class of models.
First, these models are all what may be called ``interpolating'' models.
Specifically, each is a compactification of a higher-dimensional string model $M_1$,
and as such is endowed with an adjustable compactification volume $V$.
As $V\to\infty$, we reproduce the original uncompactified string model $M_1$.
However, as $V\to 0$, we are assured by T-duality that we produce a string model
which may be considered to be the T-dual of another higher-dimensional model $M_2$.
If the compactification is untwisted, then $M_2$ will be nothing other than $M_1$.
However, if the compactification is twisted, then $M_2$ will generally differ from $M_1$.
In such cases, we can view our compactified model as smoothly ``interpolating'' 
between the uncompactified models $M_1$ (as $V\to\infty$) and $M_2$ (as $V\to 0$).
Note that the requirement that both $M_1$ and $M_2$ be {\it bona-fide}\/ self-consistent string models
provides a set of tight constraints on the twists which may be applied when 
compactifying $M_1$~\cite{BlumDienes,DienesLennekSharma,Abel:2015oxa,Aaronson}. 

The second feature that defines this class of models has to do with the choices of 
$M_1$ and $M_2$.   
Certain requirements for these choices are relatively straightforward:  for example,
we will require $M_1$ and $M_2$ to be supersymmetric and non-supersymmetric, respectively.
This guarantees that the $V\to\infty$ endpoint of the interpolation has a vanishing 
one-loop tadpole  but that our interpolating model is otherwise non-supersymmetric for all finite $V$.
This also provides
us with an ``order parameter'' $V$ for dialing the degree of supersymmetry breaking.
However, other requirements for our choices of $M_1$ and $M_2$ are less straightforward. 
In particular, for any given choice of $M_1$, only certain choices for $M_2$ 
(or equivalently only certain choices
of the SUSY-breaking twist that will be introduced into the compactification) are suitable
for generating the desired exponentially suppressed dilaton tadpole for large $V$, even if $V$ is only moderately large.
Specifically, {\it we must choose $M_2$ so that this twist leaves an equal 
number of massless bosonic and fermionic degrees of freedom in the spectrum of the 
resulting interpolating string model.}
In other words, even though this twist breaks spacetime supersymmetry (so that the resulting string spectrum contains no massless gravitinos, for example), it must be carefully chosen so that the resulting spectrum
nevertheless exhibits an equal number of massless bosonic and fermionic degrees of freedom.
Note, in particular, that there need be {\it no other relation}\/  
between the bosonic and fermionic degrees of freedom.
For example, these degrees of freedom can carry entirely different gauge charges, 
with a gluon degree of freedom balanced against
a neutrino degree of freedom.   
Likewise, some of these degrees of freedom 
can reside in a visible sector while others reside in a hidden sector.
Thus  we need not even have
equal numbers of massless bosonic and fermionic degrees of freedom in each sector separately.    
All that matters are the total numbers of massless degrees of freedom, summed over all sectors of the theory.

For any given string model $M_1$, 
it is not guaranteed that there exists a suitable model $M_2$ that will produce 
an interpolating model within this class.
In other words,  for any given model $M_1$, there may not 
necessarily exist a suitable twist that 
can be introduced upon compactification which yields a non-supersymmetric interpolating model  
with boson/fermion degeneracy at the massless level.
For this reason, 
the art of choosing suitable models $M_1$ and $M_2$ can be quite intricate, and methods for this purpose are
described in Ref.~\cite{Abel:2015oxa}.~    
But what is remarkable is 
that these are the {\it only}\/ requirements for building metastable 
string models.   
Once $M_1$ and $M_2$ are chosen satisfying these properties, a unique interpolating model is determined which
will be a member of the desired class.

\begin{figure*}[ht]
\begin{center}
  \epsfxsize 6.0 truein \epsfbox {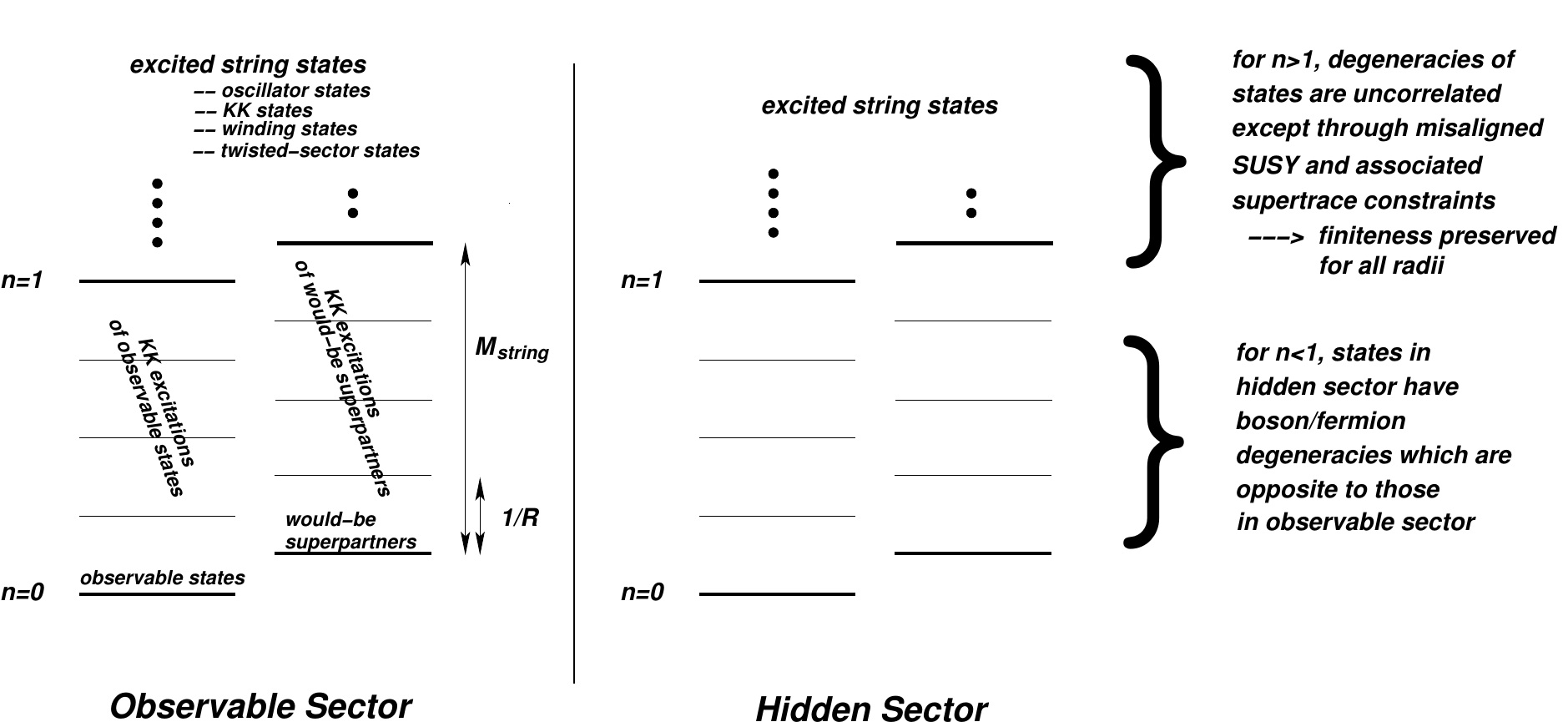}
\end{center}
\caption{The spectrum of a generic metastable interpolating model 
for $R\gg M_s^{-1}$.
States with masses below $M_{s}$ ({\it i.e.}\/, below $n=1$)  consist of massless observable states,
massless hidden-sector states,
their would-be superpartners, and their lightest KK excitations.
For these lightest states,
the net (bosonic minus fermionic) numbers of degrees of freedom
from the hidden sector are exactly equal and opposite, 
level by level,
to those from the observable sector.
This is true for all large compactification radii.
Note that this cancellation of net physical-state degeneracies between the observable and hidden sectors
bears no connection with any supersymmetry, either exact or approximate, in the string spectrum.
For the heavier states, by contrast, the observable and hidden sectors need no longer
supply equal and opposite numbers of degrees of freedom.  The properties of these
sectors are nevertheless governed by misaligned-supersymmetry constraints, as a result of which
the
entire string spectrum continues to satisfy the supertrace relations in Eq.~(\ref{straces}).
These relations maintain the finiteness of the overall string theory,
even without spacetime supersymmetry.  Figure taken from Ref.~\cite{Abel:2015oxa}.  }
\label{fig:cdcspectra}
\end{figure*}

Because the breaking of supersymmetry in this framework is tied to the compactification, 
what results is an interpolating model whose spectrum has certain characteristic features 
for large compactification volume ({\it i.e.}\/,
for $R\sim V^{1/\delta} \gg M_s^{-1}$,
where $\delta$ denotes the dimensionality of $V$ and where the symbol `$\gg$' denotes a factor of $10$ or more).
The generic spectrum of such string models is sketched in Fig.~\ref{fig:cdcspectra}. 
Situated at the massless level are 
states that together have equal numbers of bosonic and fermionic degrees of freedom.
However the would-be superpartners of these states are no longer massless,
but instead have masses $M \sim {\cal O}(1/2R)$.
This reflects the breaking of spacetime supersymmetry, leading us to a rough identification of $1/2R$ as
the scale of supersymmetry-breaking.
However, in this context it is important to stress that the massless states 
 {\it by themselves}\/ must have equal numbers of bosonic and fermionic degrees of freedom;
note in particular that this is {\it not}\/ a residual supersymmetric pairing of 
massless states with their would-be superpartners.
However, because the states with masses $M\sim {\cal O}(1/2R)$ are the would-be superpartners of the massless states,
they too will exhibit equal numbers of bosonic and fermionic degrees of freedom amongst themselves.
Note, also, that the states at each mass level may be arbitrarily split between observable and hidden sectors,
as mentioned above and indicated in Fig.~\ref{fig:cdcspectra}.~  
Consequently the equalities between the numbers of bosonic and fermionic degrees of freedom
amongst the light states
in these string models need not be observable in any way.

Proceeding further upwards in mass then leads  to a whole spectrum of repeating Kaluza-Klein (KK) excitations which
echo this basic structure,
so that the first KK excitations of the massless states have masses $M\sim {\cal O}(1/R)$
while the first KK excitations of the would-be superpartners have masses $M\sim {\cal O}(3/2R)$.
This structure is then replicated at regular mass intervals $\Delta M\sim {\cal O}(1/R)$.
Each of these levels therefore continues to exhibit equal numbers of bosonic and fermionic degrees of freedom,
even though no supersymmetry is present.
Ultimately, however, we reach the mass scale $M\sim M_s$ at which 
the first {\it string}\/ excitations appear.
In general, 
states with non-zero string excitation numbers $n>0$
have masses $M\sim \sqrt{n} M_s$.
Unlike the massless states, however, the states with $n>0$ 
need no longer come with equal numbers of bosonic and fermionic
degrees of freedom at each mass level.   Thus, for masses $M\geq M_s$, the equality between bosonic and fermionic
degrees of freedom is lost.  These states nevertheless exhibit a residual property 
called ``misaligned supersymmetry''~\cite{missusy,supertraces} 
which tightly controls the balancing between bosonic and fermionic degrees 
of freedom at {\it all}\/ mass levels throughout the infinite towers of massless and massive states, and which ensures the ultimate finiteness for which string theory is famous --- even without supersymmetry.
Indeed, misaligned SUSY is a general property of the spectra of all 
closed, tachyon-free, non-supersymmetric string  models,
and for strings within our class
guarantees that the bosonic and fermionic states are arranged 
in such a way that the ordinary supertrace relations
\beq
          {\rm Str} \, { M}^0 ~=~ 0~, ~~~~
          {\rm Str} \, { M}^2 ~\approx ~ 0~ 
\label{straces}
\eeq
nevertheless continue to hold at tree level when the 
summation is over {\it all}\/ of the physical ({\it i.e.}\/, level-matched) states in the spectrum~\cite{supertraces}.
We shall discuss the precise value of ${\rm Str}\, M^2$ for these strings below.

We are interested in this class of models because of their remarkable stability properties.
In general, for a given string model in $D$ uncompactified dimensions,
the dilaton tadpole is proportional to the one-loop vacuum amplitude (or energy density)
\begin{equation}
    \Lambda ~\equiv~ - 
            \int_{\cal F} {d^2\tau\over ({\rm Im}\, \tau)^2} \, Z(\tau,\overline{\tau})~,
\label{Lambdadef}
\end{equation}
where $Z(\tau,\overline{\tau})$ is the string partition function
and where ${\cal F}$ is the fundamental domain of the modular group.
Note that we have expressed $\Lambda$ in units of $\textstyle{1\over 2} {\cal M}^D$, 
where ${\cal M}\equiv M_s/(2\pi)$ is the reduced string scale;   thus $\Lambda$ as defined is a dimensionless
quantity, while the full energy density (cosmological constant) for the $D$-dimensional theory
is ${\cal O}(M_s^D \Lambda)$, corresponding to 
a mass scale $M_\Lambda\equiv M_s \Lambda^{1/D}$. 
However, for interpolating models within the class described above, we find that
$\Lambda$ is severely suppressed as $V\to \infty$ and indeed even for only moderately large compactification 
volumes.
For example, if we are dealing with a one-dimensional compactification ({\it i.e.}\/, a compactification
on a twisted circle) of radius $R\equiv 1/M_{\rm KK}$, 
as $M_{\rm KK}\to 0$ we find~\cite{Abel:2015oxa} 
\begin{eqnarray}
\Lambda &=&    {4 \, \Gamma[ (D+1)/2]\over \pi^{(D+1)/2}} \, [N_f^{(0)} - N_b^{(0)}]\, \left( {M_{\rm KK}\over M_s}\right)^{D}  \nonumber\\      
           && ~ +\, 
              4\,\left( {M_{\rm KK}\over M_s}\right)^{D/2} \,
\sum_{n=1}^\infty   (2\sqrt{n})^{D/2}  
                        \, [N_b^{(n)} - N_f^{(n)}]\,
             \exp\left( -4\pi \sqrt{n} M_s/M_{\rm KK} \right)~
        ~+~ ...  ~~~~
\label{Lambdaval}
\end{eqnarray}
In this expression, $D$ is the spacetime dimension of our
interpolating model (so that
$D+1$ is the spacetime dimension prior to compactification).
Likewise,
$N_{b,f}^{(n)}$ are the numbers of bosonic and fermionic degrees of freedom
in the interpolating model
at the $n^{\rm th}$ string level.
The first term in Eq.~(\ref{Lambdaval}) 
is the leading contribution from the KK excitations of the massless states,
while the remaining terms are the leading contributions from states with $n$ non-zero string excitations.
If $N_b^{(0)}\not= N_f^{(0)}$, the first term gives the leading contribution 
$\Lambda\sim M_{\rm KK}^D$, as expected.
However, as long as $N_b^{(0)}=N_f^{(0)}$, the first term vanishes and the resulting dilaton tadpole is {\it exponentially suppressed}, with a severe suppression factor of the form
$\sim \exp(-4\pi \sqrt{n} M_s/M_{\rm KK})$.  
In fact, the true suppression for $\Lambda$ is even stronger than we have indicated here since the difference $N_b^{(n)}-N_f^{(n)}$ 
tends to {\it oscillate in sign}\/ as a function of $n$.
This is ultimately a result of the misaligned supersymmetry mentioned above.
Thus the exponentially suppressed contributions from the terms with $m>0$ tend to interfere against each other, rendering the sum $\Lambda$ even more suppressed than any single term.

The result in Eq.~(\ref{Lambdaval}) is remarkable on a number of levels.
In particular, there are two aspects which are particularly surprising.
The first is the nature of the terms which can be called ``field-theoretic''.
To understand this issue, we note that
a general string theory with a compactification scale $M_{\rm KK}< M_{s}$ can be described through a sequence of
different effective theories at different energies.   For energies $E \lsim M_{\rm KK}$, the effective 
theory is a four-dimensional quantum field theory (QFT).~   Likewise, for $M_{\rm KK}\lsim E\lsim M_s$, the effective
theory is that of a higher-dimensional QFT.~   Indeed, it is only for $E\gsim M_{s}$ that our theory becomes truly stringy.  The same properties would likewise normally be reflected in the amplitudes of such a theory.
However, the construction we have described here has the remarkable property that even though $M_{\rm KK}$ is considerably below $M_s$, the single condition  
$N_b^{(0)}=N_f^{(0)}$ suffices to eliminate 
 {\it all}\/ field-theoretic contributions to 
$\Lambda$ so that $\Lambda$ depends on 
quantities such as $M_{\rm KK}$ in a completely string-theoretic
(rather than field-theoretic) manner.
This includes not only the four-dimensional QFT-like contributions to $\Lambda$,
but even the higher-dimensional QFT-like contributions.
Ultimately, this situation arises because our framework 
has the property that 
the single condition
$N_b^{(0)}=N_f^{(0)}$
actually ensures the cancellation of the net (boson minus fermion) degeneracies
at each KK level all the way up to the first non-zero string excitation. 
Thus, within our framework,
we see that the appropriately normalized $D$-dimensional energy density $M_s^D \Lambda$
receives only two groups of leading contributions
in Eq.~(\ref{Lambdaval}): 
those which scale directly as $M_{\rm KK}^{D}$
and which are therefore essentially those of a (compactified) $D$-dimensional QFT,
depending only on $N_b^{(0)}-N_f^{(0)}$,
and those which scale 
exponentially with $M_s/M_{\rm KK}$ and which are therefore intrinsically stringy,
depending on the excited string-oscillator occupation numbers $N_b^{(n)}- N_f^{(n)}$
with $n\geq 1$ in Eq.~(\ref{Lambdaval}).~ 
Indeed, the absence of other contributions which might have scaled as a higher power of $M_{\rm KK}$ and which
would have depended on the configuration of non-zero 
KK excitations below
$M_s$ is the hallmark of this framework.
Thus, enforcing the single condition $N_b^{(0)}=N_f^{(0)}$ leaves only the terms with string-theoretic suppressions
and eliminates the leading field-theoretic contributions entirely.

The second remarkable aspect of the result in 
Eq.~(\ref{Lambdaval}) concerns the severity of the exponential suppression
that arises after the 
$N_b^{(0)}=N_f^{(0)}$
condition is imposed.
Clearly, if the supersymmetry had not been broken ({\it i.e.}\/, if we had taken $M_{\rm KK}=0$), 
we would have found $\Lambda=0$.
Thus the non-zero value of $\Lambda$ indicated in Eq.~(\ref{Lambdaval}) is ultimately the result of
taking $M_{\rm KK}$ small but non-zero, so that the 
masses of the superpartner states are shifted by
an amount $\Delta M= M_{\rm KK}/2$.
Since the contribution to the string partition function $Z$ 
from a given state with mass $M$ generally scales as $e^{-M^2/M_s^2}$,
the total contribution to $\Lambda$ from any state of mass $M$ and its would-be superpartner 
of mass $M+ \Delta M$ 
can be viewed as a summation over pairwise combined contributions of the form
$e^{-M^2/M_s^2} - e^{-(M+\Delta M)^2/M_s^2}$ for various positive values of $M$.
For $\Delta M\ll M$, each such difference is approximately $(M M_{\rm KK}/ M_s^2)e^{-M/M_s}$. 
Thus, one might expect that the mass shifts between the states in our 
theory and their would-be superpartners would generate
a total contribution to $\Lambda$ 
which is suppressed as a {\it power}\/ of $M_{\rm KK}/M_s$.
However, the set of masses $M$ over which such a summation is performed is itself dependent on $M_{\rm KK}$
and becomes dense as $M_{\rm KK}/M_s \to 0$.
Thus, as $M_{\rm KK}\to 0$, the cancellation between states  
and their would-be superpartners becomes more complete while the {\it density}\/ of such states increases.
It is ultimately the interplay between these two effects --- along with our condition $N_b^{(0)}=N_f^{(0)}$ 
--- which produces the severe ``inverted'' suppression
factor $e^{-M_s/M_{\rm KK}}$ quoted in Eq.~(\ref{Lambdaval}).

What we obtain, then, is a four-dimensional non-supersymmetric string theory 
governed by three fundamentally different mass scales.
The first is $M_{\rm KK}$, which governs the splitting between states and their would-be superpartners
and which may thus be viewed as 
the scale of supersymmetry breaking.
The second is $M_s$, which governs the energies associated with the string oscillator excitations and which 
therefore serves as the scale of the UV completion of the theory.
Remarkably, however, when $N_b^{(0)}=N_f^{(0)}$, these two scales conspire to produce 
a new scale   
\beq
       M_\Lambda ~\equiv~ \sqrt{ M_{\rm KK} M_s} \, e^{-\pi M_s/M_{\rm KK}}
\label{MLambda}
\eeq
which is significantly smaller than either of the
two previous scales and which sets the magnitude of the  corresponding one-loop 
cosmological constant (vacuum energy).
It is this scale which governs the ultimate tree-level dilaton stability of the theory.

The suppression of this last scale can also be understood geometrically.
Because supersymmetry is broken through compactification in our construction,
massive string modes need to propagate over the full compactification volume, {\it i.e.}\/, over a distance $2\pi R$,
in order to realize a non-zero $\Lambda$.
This leads to a Yukawa suppression of the form $\exp(-2\pi RM_s)$.
This geometric understanding ties in 
with our alternative explanation above since 
the inverted suppression factor and the ``large-volume'' Yukawa picture
both arise after Poisson resummation.

Viewed from the perspective of 
string model-building, however,
the result in Eq.~(\ref{Lambdaval}) is extremely beneficial.
As we have already noted, the scale of supersymmetry breaking in this construction
can roughly be taken to be ${\cal O}(M_{\rm KK}/2)$, 
or ${\cal O} (1/2R)$ in the one-dimensional case.
However, as long as we ensure that $N_b^{(0)}=N_f^{(0)}$, 
the dilaton-stability of such a string (\ie, the 
suppression of the corresponding value of $\Lambda$) is not polynomial in $M_{\rm KK}$ but {\it exponential}\/.
We can therefore dial $M_{\rm KK}$ (or more generally our compactification volume $V$) 
to any value desired  --- even to the TeV scale --- while nevertheless maintaining the 
required suppression of the dilaton tadpole and assuring the metastability of the non-supersymmetric string.
It is for this fundamental reason that our framework leads to a promising starting point for a non-supersymmetric
string phenomenology.
Furthermore,
the value of the supertrace 
${\rm Str}\, M^2$ 
for any tachyon-free closed string theory compactified to four dimensions
can be shown~\cite{supertraces} to scale as $M_s^{2} \Lambda$, where $\Lambda$ is defined
as in Eq.~(\ref{Lambdadef}).
Thus, the severe suppression of $\Lambda$ for all string models in this class
additionally becomes a suppression for ${\rm Str}\, M^2$: 
\beq
 {\rm Str}\, M^2 ~\sim~ {\cal O} (M_\Lambda^4/M_s^2) ~\sim~ {\cal O}(M_{\rm KK}^2\, e^{-4\pi M_s/M_{\rm KK}})~.
\eeq
Again, we stress that this occurs {\it even though the scale of SUSY-breaking in this framework 
is ${\cal O}(M_{\rm KK}/2)$}\/.

When in the following we construct explicit string models,
we shall focus on a specific configuration 
within this general framework
which forms a particularly useful testing ground for the more general discussion.  
As we shall discuss,
this configuration
is based on perturbative ten-dimensional heterotic strings exhibiting large (GUT-like) gauge symmetries
and proceeds through two stages of compactification.
The first is a compactification down to $4+\delta$ dimensions
on a manifold or orbifold $\mathbb K$  with volume
$R\sim M_s^{-1}$
such that the resulting $(4+\delta)$-dimensional string model is supersymmetric. 
By contrast, 
the second stage of compactification from $4+\delta$ to
four dimensions 
occurs through a $\delta$-dimensional compactification/interpolation of the type we have been discussing.
The space on which 
this compactification occurs is a 
$\delta$-dimensional freely-acting orbifold $\mathbb{O}_\delta$.
We thus have

\bigskip
\begin{center}
   \fbox{ \begin{minipage}{0.9 truein}
    \begin{center}
     10D ${\cal N}$=1~SUSY\\ 
      GUT-like\\ string model 
    \end{center}
   \end{minipage}}
    $~~\xrightarrow[R\sim M_s^{-1}]{\mathbb{K}}~~$
   \fbox{ \begin{minipage}{1.15 truein}
    \begin{center}
     (4+$\delta$)-dimensional\\ ${\cal N}$=1~SUSY\\  
      GUT-like\\ string model 
     \end{center}
   \end{minipage}}
    $~~\xrightarrow[R\gg  M_s^{-1}]{\mathbb{O}_\delta}~~$
   \fbox{ \begin{minipage}{1.55 truein}
    \begin{center}
     4D non-SUSY\\ massless B/F-degeneracy\\ SM-like\\ string model 
     \end{center}
   \end{minipage}}
\end{center}
\bigskip

\noindent In Ref.~\cite{Abel:2015oxa} and in our models to be presented below,
we take $\delta=2$ (so that our intermediate model is six-dimensional) and  
$\mathbb{O}_{\delta} \equiv \mathbb{T}_{2}/\mathbb{Z}_{2}$ 
(with the understanding that the $\mathbb{Z}_{2}$ acts on both $\mathbb{T}_{2}$ and $\mathbb K$). 
If this orbifold were untwisted, we would obtain an ${\cal N}=1$  supersymmetric 
theory in four dimensions.
However, we introduce a Scherk-Schwarz twist 
which acts not only on the spacetime degrees of freedom but also on the internal gauge degrees of freedom.
This coupling between the spacetime twist and the internal gauge twist is ultimately required by 
modular invariance. 
We also choose these 
twists so as to additionally satisfy the conditions laid out above, including the requirement of bose/fermi degeneracy
at the massless level.
This then produces a non-supersymmetric four-dimensional 
string model with the desired metastability properties.

In this configuration,
both the original GUT-like gauge symmetry and the original spacetime supersymmetry 
are broken together in the final stage of compactification.
It is this feature which ultimately leads to the GUT-precursor 
and entwined-supersymmetry structures 
which are the focus of this paper.
In fact, we shall even eventually argue in Sect.~\ref{discussion} 
that these structures transcend our particular
string construction, and are inevitable within broad classes
of non-supersymmetric UV-complete theories.
It is therefore to these topics that we now turn.

\section{GUT precursors and entwined SUSY
\label{entwined}}

As indicated above, the starting point of our construction is
a higher-dimensional string model exhibiting not only spacetime SUSY but also a GUT gauge symmetry.
Both of these symmetries are then broken together upon the final stage of compactification.
In principle, the exponential suppression of the dilaton 
tadpole does not require that we begin with a GUT symmetry prior to compactification.
Nor does it require that 
this symmetry be broken by compactification.
Ultimately, these additional features are needed for phenomenological purposes.
In this section, we shall begin by explaining why these additional features 
are needed.
We shall then demonstrate that these features inevitably lead to a GUT-precursor
structure and the emergence of an entwined supersymmetry.
As we shall see, these phenomenological aspects are both quite general
and can be understood from a geometric point of view.
Indeed, as we shall demonstrate, both entwined SUSY as well as the
GUT-precursor structure are rather generic phenomenological properties of a wide class of 
non-supersymmetric strings --- even independently of the need for 
boson/fermion degeneracy of the massless states.
This section thus constitutes the main theoretical 
portion of this paper, 
with subsequent sections providing explicit constructions that illustrate these assertions.

\subsection{The problem of gauge couplings in large-volume compactifications\label{2a}}

Because our dilaton-stabilization mechanism requires the existence of a large compactification 
volume, an immediate problem that emerges concerns the values of the gauge and gravitational couplings
(sometimes referred to as the ``decompactification problem'').
It is easy to see how this problem arises.
For concreteness, let us consider the case of the heterotic string compactified from ten to four dimensions.
In general, the coupling expansion for an $n$-point genus-$g$ diagram behaves as 
\begin{equation}
  V_{6}\, g_{10}^{n-\chi}~,
\label{eq:1}
\end{equation}
where $g_{10}$ is the ten-dimensional string coupling, 
where $V_6$ is the compactification volume, 
and where $\chi\equiv 2(1-g)$  
is the (topologically invariant) Euler number of the string worldsheet.
Thus at tree level ({\it i.e.}, for $g=n=0$) the  effective four-dimensional Lagrangian for gravitational and gauge interactions
scales as $1/g_{10}^2$
and takes the form
\begin{equation}
S~=~\int d^{4}x\, \frac{V_{6}}{g_{10}^{2}}\, \left(\alpha'^{-4}{\cal R}+\alpha'^{-3}F^{2}\right)~,
\end{equation}
where ${\cal R}$ is the Ricci scalar and $F$ is a gauge field strength.
From this we can read off the effective four-dimensional tree-level gauge coupling $g_{\rm 4}$ and effective
four-dimensional Planck scale $M_P$:
\begin{equation}
      {1\over g_{\rm 4}^2} ~=~ {v_6\over g_{10}^2}~,~~~~~~~
      M_P^2 ~=~ {v_6\over g_{10}^2 \alpha'}~,
\label{relations}
\end{equation}
where $v_{6}=V_{6}/(\alpha')^3$ is the compactification volume normalized with respect to the fundamental string scale.
From these results it follows that $\alpha' M_P^{2} = g_{\rm 4}^{-2}$,
or equivalently $M_s= g_{\rm 4}M_P$.

The relations in Eq.~(\ref{relations}) are completely general.
Moreover, for compactification volumes near the string scale [{\it i.e.}\/, for $v_6\sim {\cal O}(1)$],
we find that $g_{\rm 4}\sim g_{10}$.
The perturbativity condition $g_{10}\lsim 1$ then requires $g_{\rm 4}\lsim 1$,
and one often chooses $g_{\rm 4}\sim g_{\rm GUT} \approx 1/\sqrt{2}$ 
in order to make contact with standard logarithmic gauge coupling unification. 
Indeed, in such a scenario,
the measured four-dimensional SM gauge couplings at the weak scale run logarithmically up to the GUT/string
scale where they unify into $g_{\rm GUT}$.   Note that these   gauge couplings run logarithmically
over this range precisely because $v_6\sim {\cal O}(1)$, so that the theory is effectively
four-dimensional below $M_s$.
Thus, for $v_6 \sim {\cal O}(1)$,  the usual logarithmic gauge coupling unification 
can be preserved and naturally embedded into string theory~\cite{Dienespath}.

The situation is very different when the compactification volume is large, as in our configuration.
In this case $v_6\gg 1$, whereupon  the perturbativity condition $g_{10}\lsim 1$
implies $g_{\rm 4}\ll 1$.
Such small values for $g_{\rm 4}$ are difficult to reconcile with the 
measured values of the four-dimensional gauge couplings at the electroweak scale.
Of course, the assumption of a large compactification volume $v_6\gg 1$ 
implies that $M_{\rm KK}\ll M_s$, so 
that our theory is actually higher-dimensional between $M_{\rm KK}\sim V_6^{-1/6}$ and $M_s$.
This then opens up an interval over which
the running of the gauge couplings above $M_{\rm KK}$ has a power-law (rather than logarithmic) 
dependence.
However, even this observation cannot evade our difficulties.
First, with power-law running above $M_{\rm KK}$, the traditional logarithmic gauge coupling unification is generally lost. 
Moreover, even though power-law running can still produce a power-law unification of the gauge 
couplings as one proceeds upwards in energy~\cite{DDG1,DDG2,DDG3}, this
unification typically occurs very rapidly after the onset of KK modes, with
$M_{\rm GUT}/M_{\rm KK}$ never very large.
Identifying $g_{\rm GUT}\sim g_{\rm 4}$ and $M_{\rm GUT}\sim M_s$,
we see that this therefore does not leave much room 
for a large compactification volume $v_6$.
Or, phrased somewhat differently, we might continue to insist that $v_6\sim (M_s/M_{\rm KK})^6 \gg 1$,
but this would no longer permit us to parametrically identify $M_{\rm GUT}$ with $M_s$ --- a feature
that we would generally like to retain.  

The question then arises as to how we can reconcile the measured ${\cal O}(1)$ values of
the four-dimensional gauge couplings at low energies
with an ${\cal O}(1)$ value of the ten-dimensional string coupling $g_{10}$ --- all in the presence
of a large compactification volume $v_6\gg 1$, and all while
preserving a logarithmic gauge coupling unification at $M_{\rm GUT}\sim M_s$.

\subsection{GUT precursors\label{precurs}}

It turns out that all of these features are not only reconciled but also realized 
naturally within the so-called ``GUT precursor'' scenario 
originally presented in Refs.~\cite{Dienes:2002bg,Dienes:2004rt}.
The discussion in Refs.~\cite{Dienes:2002bg,Dienes:2004rt} was essentially field-theoretic,
but we shall see that this scenario 
is also a natural prediction of our string framework.

The basic thrust of the scenario presented in Refs.~\cite{Dienes:2002bg,Dienes:2004rt}
is to develop a self-consistent understanding of gauge coupling unification in the presence of large extra spacetime dimensions.    
For simplicity,
let us 
imagine a $(4+\delta)$-dimensional theory exhibiting a grand-unified symmetry $G_{\rm GUT}$.
Let us furthermore imagine breaking this symmetry down to the Standard-Model (SM) gauge group
through an orbifold compactification of the $\delta$ extra dimensions.
For simplicity, we shall imagine that each of
these extra dimensions is compactified on a circle with radius $R$,
along with an overall orbifold twist which is designed 
not only to preserve the zero modes of those gauge fields which survive the GUT symmetry breaking
(such as the gauge bosons of our SM gauge group), but also to project out the zero modes
of those remaining gauge fields (such as the $X$  and $Y$ gauge bosons) which are exotic from 
the point of view of the SM gauge group but which were otherwise needed in order to fill out $G_{\rm GUT}$.
Thus, at low energies, our spectrum consists of only the SM zero modes, and the original GUT symmetry appears broken.
Indeed, the lowest-lying exotic states are the $X$ and $Y$ gauge bosons which do not appear in the resulting 
spectrum until the first excited KK level, with masses $\sim 1/R$.
Of course, 
the full grand unification 
does not occur until the low-energy gauge couplings actually unify at some much higher scale $M_{\rm GUT}$.
Thus, we immediately observe a remarkable feature of GUT breaking by orbifolds (as opposed to, say, the
more traditional GUT breaking via
a Higgs mechanism):   although the actual grand unification (as evidenced through the unification of gauge couplings) only occurs at $M_{\rm GUT}$,
the first experimental signatures (or ``precursors'')
of the impending unification are the $X$  and $Y$ gauge boson
states which first appear at $M\sim 1/R$ --- a scale which is parametrically distinct from $M_{\rm GUT}$.
The question then arises as to how large a separation of scales can be tolerated between the precursor scale $1/R$
and the unification scale $M_{\rm GUT}$.   
In other words, how large a compactification volume
can be tolerated in such a scenario?   What is the maximum allowed value of $M_{\rm GUT}R$?

This is the question addressed in Refs.~\cite{Dienes:2002bg,Dienes:2004rt}.   
Remarkably, what was found is that $M_{\rm GUT} R$ can actually grow {\it arbitrarily large}\/.
The criteria leading to this possibility can be understood as follows.
In the presence of $\delta$ extra spacetime dimensions of radius $R$, the low-energy gauge couplings
$\alpha_i$ (as measured, say, at $M_Z$) evolve upwards in energy (to an arbitrary high scale $\Lambda$) 
according to the approximate one-loop RGE's~\cite{TV,DDG1,DDG2,DDG3}
\beq
\alpha_i^{-1}(\Lambda) ~\approx~ \alpha_i^{-1}(M_Z) - {b_i\over 2\pi} \ln {\Lambda\over M_Z} + 
                                                {\tilde b_i\over 2\pi} \ln \Lambda R 
  - {\tilde b_i X_\delta\over 2\pi\delta} \left[ (\Lambda R)^\delta -1\right]~,
\label{RGE}
\eeq
where $b_i$ are the beta-function coefficients of the zero-mode fields,
where $\tilde b_i$ are the beta-function coefficients associated with the field content at each excited KK level,
and where $X_\delta\equiv 
\pi^{\delta/2}/\Gamma(1+\delta/2)$
is the volume of the unit ball in $\delta$ dimensions.
It is the presence of KK states running in the loops that causes the evolution to follow a power-law behavior.
As we shall shortly see, this generic form is also borne out by explicit string results within the particular
six-dimensional configuration discussed at the end of Sect.~\ref{sec:framework}.

In a scenario with arbitrary values of $\tilde b_i$, each low-energy gauge coupling experiences an independent
power-law evolution and the measured low-energy couplings are grossly inconsistent with unification.  
However, there do exist values of $\tilde b_i$ for which our low-energy gauge couplings experience not only
power-law evolution but also unification~\cite{DDG1,DDG2,DDG3}.   One example is an accelerated, power-law evolution which usually occurs soon after the onset of the KK modes, leading to values of $M_{\rm GUT}R$ which are tightly constrained and often smaller than a single order of magnitude.

There is, however, a second option~\cite{orbifoldguts}:  all $\tilde b_i$ can be equal, 
with $\tilde b_i=\tilde b$ for all $i$.
In this case each gauge coupling continues to experience a power-law running, but the {\it differences}\/ between
 the gauge couplings evolve only logarithmically.  Indeed, for appropriate values of $b_i$, we can reproduce
a logarithmic unification which inevitably occurs at the traditional high scale $M_{\rm GUT}$.    
Of course, since each individual coupling experiences a power-law evolution over this entire energy range,
we must be sure that none of these couplings hits a Landau pole en route to unification or 
otherwise accrues a value which
would invalidate our overall implicit perturbativity assumptions.   This generally requires that $\tilde b<0$.   
This in turn ensures that our measured individual gauge couplings at low energies flow to extremely small (rather than extremely large) values in the UV, ultimately yielding  a unified gauge coupling $g_{\rm GUT} \ll 1$.
Thus, in this manner, the very small values 
$g_{\rm GUT}\sim g_{\rm 4}\ll 1$ can naturally be reconciled with the measured ${\cal O}(1)$ values of the
gauge couplings at low energies, all while preserving a traditional {\it logarithmic}\/ unification of gauge couplings
and a correspondingly large value for $M_{\rm GUT}R$.   Hence gauge coupling unification survives, even with large-volume compactifications.

There is also another way to understand this result and to verify its perturbativity.
In theories such as this for which there are many degrees of freedom,
an effective measure for the strength of gauge interactions is not the gauge coupling $\alpha_i$ but
rather the 't~Hooft coupling $\tilde \alpha_i \equiv N\alpha_i$, where $N$ is a measure of the 
number of degrees of freedom running in the loops. Indeed, for any energy scale $\Lambda$, we may take $N$ as the
number of KK levels that have already been crossed, {\it i.e.}\/, 
$N \equiv X_\delta (\Lambda R)^\delta$.   
According to Eq.~(\ref{RGE}),
the individual gauge couplings $\alpha_i$ all scale in the UV (\ie, for $\Lambda R\gg 1$) as $\alpha (\Lambda)\approx -2\pi\delta (\Lambda R)^{-\delta}/(\tilde b X_\delta)$. Thus the corresponding 't~Hooft couplings
scale as $\tilde \alpha \approx -2\pi\delta/\tilde b$. In other words,
as originally noted in Ref.~\cite{agashe},
the effective 't~Hooft couplings $\tilde \alpha_i$ become independent of $\Lambda R$
as $\Lambda R$ increases 
and actually approach a  {\it UV fixed point}\/ $\tilde \alpha \approx -2\pi\delta/\tilde b$. 
Moreover, this UV fixed point is perturbative so long as $\tilde \alpha \ll 4\pi$, or
$\delta /(2\tilde b) \ll 1$
--- indeed, the 't~Hooft coupling $\tilde \alpha$ can then be interpreted
as the dimensionless coupling associated with the $(4+\delta)$-dimensional theory
that emerges in the infinite-volume limit. 
Consequently, if $\tilde b$ is sufficiently large and negative, there is no obstruction to having an 
arbitrarily large compactification volume with $\Lambda R\gg 1$.
This is the underlying reason why this scenario can tolerate 
a large separation of scales between the GUT precursor scale $1/R$ and
the unification scale $M_{\rm GUT}$.

It is not difficult to realize such theories in a natural way.   For example, let us imagine, as in 
Ref.~\cite{Dienes:2002bg}, that our zero-mode
fields exhibit ${\cal N}=1$~SUSY and are those of the MSSM, 
while our  unified gauge group is $SU(5)$.  Let us further imagine that only one extra dimension is compactified,
{\it i.e.}\/, $\delta=1$.   It then follows that the states at each excited KK level 
are ${\cal N}=2$~SUSY vector multiplets transforming in the adjoint of $SU(5)$, 
with $\tilde b_i=\tilde b= -10$ for all $i$.   
This then leads to a unified perturbative fixed-point coupling $\tilde \alpha\approx 0.63$.

The presence of ${\cal N}=2$ multiplets at each excited KK level is an extremely beneficial outcome,
since the presence of ${\cal N}=2$~SUSY in the bulk ensures that any higher-loop power-law effects are suppressed
by a factor of $1/(\Lambda R)$ relative to the one-loop effects.
Such higher-loop effects therefore become increasingly 
insignificant for $\Lambda R\gg 1$~\cite{DDG1,DDG2,Kakushadze:1999bb,Dienes:2002bg,Dienes:2004rt}.
Likewise, there can be other effects (such as non-universal logarithms or 
contributions from brane-kinetic terms~\cite{Dienes:2002bg}) which, at first glance,
also appear to have the power to eliminate the logarithmic unification in this scenario.
However it can be shown~\cite{Dienes:2002bg} that such effects are ultimately subleading and
generally leave the unification intact.

Thus far, we have shown how the measured low-energy gauge couplings $\alpha_i(M_Z)$ can, through power-law
running associated with a large compactification volume, lead to a logarithmic gauge coupling unification
$\alpha_i\approx \alpha_{\rm GUT}$ at a relatively high scale $M_{\rm GUT}$.    As we have seen,
the principal required ingredients are 
the existence of complete GUT 
multiplets at each excited KK level,
the presence of ${\cal N}\geq 2$ SUSY at each excited KK level,
and a field content at each excited KK level such that 
$\tilde b<0$.
These properties for the excited KK states ensure that the differences between the
low-energy couplings $\alpha_i(\Lambda)$ 
evolve at most logarithmically,
that each individual coupling becomes extremely weak at the GUT scale for $\Lambda R\gg 1$,
and that the contributions from higher loops do not disturb our one-loop results.
Properly choosing the field content of the zero modes then ensures that these couplings actually unify,
just as they would have in four dimensions.

Given this field-theoretic scenario,
the final step is to embed this scenario within a UV-complete theory
such as string theory.   However, this is not difficult to arrange:
we simply identify $\alpha_{\rm GUT}$ with the four-dimensional gauge coupling $\alpha_4$ 
in Eq.~(\ref{relations}).
Likewise, we identify $M_{\rm GUT}$ with $M_s$.
Note that our identification of $M_{\rm GUT}$ with $M_s$ is not meant to be a precise one, for there can be many 
${\cal O}(1)$ effects which could explain a discrepancy between $M_{\rm GUT}$ and $M_s$.
Such effects are reviewed, for example, in Ref.~\cite{Dienespath}.
Likewise, at first glance it may seem strange to match a one-loop ``bottom-up'' coupling such as 
$\alpha_{\rm GUT}$ with a tree-level
``top-down'' string coupling such as $\alpha_4$.  
However, this lopsided matching between a one-loop coupling and a tree-level coupling arises
only because our determination of $\alpha_4$ was itself
the result of a tree-level string analysis. 
Indeed, we could equally well have performed a more complete one-loop string analysis, 
carefully integrating out all heavy string states before applying our matching conditions.
In such a case, we would then 
understand the volume dependence as arising from string threshold corrections. 
As an example, in the toroidal 6D case (such as in the explicit example we shall present later),
the result is expressed in terms of the usual moduli for the ${\mathbb T}_2$ compactification, namely 
$T$ and $U$.
The gauge couplings are then found to behave universally as 
\begin{equation}
\alpha_i^{-1}(\mu) ~=~ \alpha_i^{-1}(M_s) +\frac{b_i}{2\pi}\ln\frac{M_{s}}{\mu}+\frac{\Delta}{4\pi} \, ,
\label{eq:flow}
\end{equation}
where~\cite{Angelantonj:2015nfa,Abel:2016hgy} 
\begin{eqnarray}
\Delta_{} & ~=~ & -\tilde{b}_{}\log\left(4T_{2}\tilde{U}_{2}|\eta(iT)|^{4}|\eta(i\tilde{U})|^{4}\right)+(\tilde{b}_{}-b_{i})\log\left(4T_{2}\tilde{U}_{2}|\vartheta_{4}(iT)|^{4}|\vartheta_{4}(i\tilde{U})|^{4}\right)\,,\nonumber \\
 & ~=~ & \frac{\pi}{3}\tilde{b}_{}\left(T_{2}+\tilde{U}_{2}\right)-b_{i}\log\left(4T_{2}\tilde{U}_{2}\right)+\mathcal{O}(e^{-\pi\tilde{U}_{2}},e^{-\pi T_{2}})~.
\label{eq:largevolD}
\end{eqnarray}
Here $iT=T_1+iT_2$ and $iU=U_1+iU_2$ with $i\tilde{U}=-1/(iU-1)$, where $T_2$ is the compactification two-volume. 
In Eq.~(\ref{eq:largevolD})
we see both the $\tilde{b} v_d$ leading terms and the logarithmic contribution from the running of the ${\cal N}=1$ sector between the KK scale and the string scale.

Given that we are forced to embed the GUT-precursor structure into string theory,
one natural issue is to determine the scale at which gravitational effects become strong.
Since $M_s= g_4 M_P$ and $g_4\ll 1$, it follows that $M_P\gg M_s$.
At first glance, this might seem to imply that gravitational effects do not arise
until far beyond the string scale.   However this scenario involves a large
volume of compactification, and it is well known that under such circumstances
the actual quantum-gravity scale $M_\ast$ is given by $M_\ast\sim (M_P^2/V_\delta)^{1/(2+\delta)}$
where $V_\delta\sim R^\delta$ is the volume of compactification.
We thus consistently find that $M_\ast\sim M_s$, implying that the effective quantum-gravity scale
is not greatly separated from the string scale, just as occurs in more traditional scenarios involving
Planck-scale compactification volumes. 

Combining the above relations, we find that
\beq
            (M_s R)^{\delta/2} ~=~ \sqrt{ {-8\pi^2 \delta \over \tilde b X_\delta}} \, {M_P\over M_s}~.
\label{scalerelation}
\eeq
Thus the Planck scale $M_P$, the string scale $M_s$, and the
Kaluza-Klein scale $M_{\rm KK}\equiv R^{-1}$ are all balanced together 
in any self-consistent heterotic string-theoretic scenario.
It is useful to examine some representative cases.
If $M_P\approx 10^{18}~{\rm GeV}$ and
we identify $M_s\approx M_{\rm GUT}\approx 10^{16}~{\rm GeV}$, we find
$(M_s R)^{\delta/2}\approx 10^2$.   For $\delta =1$, this implies  $R^{-1}\approx 10^{12}$~GeV,
while for $\delta=2$ this implies $R^{-1}\approx 10^{14}$~GeV.
Indeed, taking larger values of $\delta$ only increases the value of $R^{-1}$.
From the perspective of the low-energy theory, this is an extremely large scale for SUSY-breaking.
We nevertheless find from Eq.~(\ref{Lambdaval}) that 
$\Lambda \sim \exp\left( -4\pi 10^4\right)$ for $\delta=1$ 
and  
$\Lambda \sim \exp\left( -4\pi 10^2\right) $ for $\delta=2$,
assuming that $N_b^{(0)}=N_f^{(0)}$ in each case.
Thus the dangerous one-loop dilaton tadpole is extremely suppressed and essentially
zero for all practical purposes.

It is important to note that the overall scaling relations we have been working with are ultimately
governed by $\tilde b$, the universal beta-function coefficient associated with the matter content
of the excited KK states.  
By contrast, the value of the unification scale $M_{\rm GUT}$ is set by the values of the 
individual beta-function coefficients $b_i$ associated with the zero-mode states.
Thus, there remains the freedom --- just as in all field-theoretic GUT scenarios --- 
to choose our low-lying matter content in such a way as
to alter these beta-function coefficients and thereby adjust the unification scale.
In this way, it might even be possible to bring $M_{\rm GUT}$ significantly below the traditional
unification scale.
Continuing to identify $M_s$ with $M_{\rm GUT}$ would then lead to a self-consistent scenario in which
$M_s \ll 10^{16}$~GeV, with $R^{-1}$ correspondingly reduced even further,
perhaps even all the way into the TeV range.   
Indeed, taking $R^{-1}\approx 1$~TeV within Eq.~(\ref{scalerelation}),
we find that $M_s\approx 10^{13}$~GeV for $\delta=1$,
whereupon we see that $M_s/M_{\rm KK}\approx 10^{10}$.
Even for $\delta=6$ we find $M_s\approx 10^6$~GeV, whereupon  $M_s/M_{\rm KK}\approx 10^3$.
Thus, even in such cases with significantly reduced string scales, we continue to find 
that the one-loop dilaton tadpole is extremely suppressed.
Such scenarios thus retain dilaton stability
and incorporate not only an effective TeV-scale breaking of SUSY
but also GUT-precursor states that are potentially observable at the TeV scale --- orders of magnitude lower
than the scale of gauge coupling unification!   
Such states would have the gauge quantum numbers of leptoquarks and would thus give 
rise to dramatic collider signatures.

\subsection{The structure and ubiquity of $e$-SUSY}
\label{eSUSYintro}

As we have seen, 
non-supersymmetric strings can be made stable in a consistent fashion
if the excited KK modes consist of GUT representations
falling into ${\cal N}\geq 2$ supermultiplets.
Therefore, in order to obtain the 
Standard Model for the zero modes, our compactification must not only break the GUT gauge symmetry but 
also simultaneously break the remaining supersymmetry.   As we now discuss,
this inevitably gives rise in the resulting theory to a structure involving what we call ``entwined SUSY'' ($e$-SUSY).
We begin by describing more explicitly what we mean by entwined-SUSY, both in terms of the
allowed spectrum as well as the allowed couplings. We shall
then discuss why this structure arises.

As an example, let us consider a theory in which an underlying $SU(5)$ ${\cal N}=1$ GUT model contains two generations of chiral supermultiplets,
${\overline{{\bf 5}}}_{0}$ and ${\overline{{\bf 5}}}_{\frac{1}{2}}$. 
Here the subscripts indicate the charges under a horizontal $U(1)$ symmetry which we will refer to generically as $Q_{\rm horiz}$.
(These multiplets may also carry other horizontal charges, but only one horizontal charge is needed in order
to illustrate the entwined-supersymmetric structure.)  
Under the group decomposition $SU(5)\to SU(3)_c\times SU(2)_L$,
we recall that
${\overline{{\bf 5}}}=({\overline{{\bf 3}}},1) \oplus \,(1,  {\overline{\bf 2}})$.
Our two chiral multiplets
${\overline{{\bf 5}}}_{0}$ and ${\overline{{\bf 5}}}_{\frac{1}{2}}$ 
thus have states consisting of
\begin{eqnarray}
{\overline{\bf 5}}_0  &=&
\left\{
\begin{array}{rrl}
{\mbox {fermions}}: & ({\overline{{\bf 3}}} ,1)_{0}\oplus (1, {\overline{\bf 2}})_{0} ~&\equiv ~ d^c_{0} \oplus  \ell_{0} \\
{\mbox {bosons}}: & ({\overline{{\bf 3}}} ,1)_{0}\oplus (1, {\overline{\bf 2}})_{0} ~&\equiv ~ \tilde d^c_{0} \oplus  \tilde\ell_{0}
\end{array} \right.  \nonumber\\
{\overline{\bf 5}}_{{1\over 2}} &=&
\left\{
\begin{array}{rrl}
{\mbox {fermions}}: & ({\overline{{\bf 3}}} ,1)_{{1\over 2}}\oplus (1, {\overline{\bf 2}})_{{1\over 2}} ~&\equiv ~ d^c_{{1\over 2}} \oplus  \ell_{{1\over 2}} \\
{\mbox {bosons}}: & ({\overline{{\bf 3}}} ,1)_{{1\over 2}}\oplus (1, {\overline{\bf 2}})_{{1\over 2}} ~&\equiv ~ \tilde d^c_{{1\over 2}} \oplus  \tilde \ell_{{1\over 2}}~,
\end{array} \right.
\label{5proja}
\end{eqnarray}
where the hypercharges (which we do not show) are the canonical ones.
Indeed, this is the matter content that would emerge upon compactification without the crucial Scherk-Schwarz twists.

Implementing the twists then eliminates part of this matter content.
Of course, which states survive and which are projected out depends on the details 
of the relevant Scherk-Schwarz twists and GSO projections.
In many simple string constructions, these projections would eliminate either one or the other of these supermultiplets.  Likewise, supersymmetry would be broken if the internal structure of each multiplet was also destroyed,
leaving behind bosonic and fermionic states that could no longer be paired with each other.
However, what we find 
in the configuration described above  --- where the last stage of compactification breaks the supersymmetry
and GUT symmetry simultaneously --- 
is that the projections 
instead lift the masses of certain states 
according to a non-trivial combination of their SM representations, horizontal charges and spin-statistics.
Indeed, what remains at the massless level are a set of states
which together fill out
a single light ``fake'' supermultiplet  which we shall denote ${\overline{\bf 5}}_e$:
 \begin{equation}
{\overline{\bf 5}}_e\,\, =\,\,
\left\{
\begin{array}{rrl}
{\mbox {fermions}}: & ({\overline{{\bf 3}}} ,1)_{0}\oplus (1, {\overline{\bf 2}})_{\frac{1}{2}} ~&\equiv ~ d^c_{0} \oplus  \ell_{\frac{1}{2}} \\
{\mbox {bosons}}: &  ({\overline{{\bf 3}}} ,1)_{\frac{1}{2}}\oplus (1, {\overline{\bf 2}})_{0} ~&\equiv ~ {\tilde{d}}^c_{\frac{1}{2}} \oplus  \tilde{\ell}_{0} ~ .
\end{array}\right.
 \label{5proj}
\end{equation}
All other components from the original pair of  ${\overline{{\bf 5}}}$'s in Eq.~(\ref{5proja}) are
given masses of order the compactification scale.

It is immediately apparent from the structure of the multiplet in Eq.~(\ref{5proj})
that the SUSY-breaking is ``entwined'' with the horizontal charges in a non-trivial way.
Indeed, the matter spectrum is not symmetric under a supersymmetry transformation alone,
but only a supersymmetry transformation coupled with a permutation of $Q_{\rm horiz}$ charges.
Thus, the supersymmetry is completely broken in the resulting theory.
For example, no massless gravitinos or gauginos survive in the massless spectrum.
Nevertheless, due to the controlled structure of the SUSY-breaking,
a residual imprint of the original supersymmetry remains.
This is our 
``entwined'' SUSY ({\it e}\/-SUSY).

A similar entwining also occurs for the other GUT multiplets.
For example,
suppose  that ({\it e.g.}\/, as dictated by anomaly cancellation) the content of our
original GUT model fills out entire SM generations with the  inclusion of  a 
${{\bf 10}}_{0}$ representation and a ${{\bf 10}}_{\frac{1}{2}}$ representation.
In the spontaneously broken theory,
the entwined {\bf 10}\/ multiplet is then given by
\begin{equation}
{ {\bf 10}}_e\,\, =\,\,
\left\{
\begin{array}{rrl}
{\mbox {fermions}}: & ( {{\bf 3}} ,{{\bf 2}})_{\frac{1}{2}}\oplus (\overline {{\bf 3}},1 )_{0}\oplus (1,1 )_{0} ~& \equiv ~ q_{\frac{1}{2}} ~\oplus u^c_{0} \oplus  e^c_{0}  \\
{\mbox {bosons}}: & ( {{\bf 3}} ,{{\bf 2}})_{0}\oplus (\overline {{\bf 3}},1 )_{\frac{1}{2}}\oplus (1,1 )_{\frac{1}{2}}~& \equiv ~ \tilde{q}_{0} \oplus \tilde{u}^c_{\frac{1}{2}}~ \oplus  \tilde{e}^c_{\frac{1}{2}} ~ , 
\end{array}\right.
\label{10proj}
\end{equation}
where we have adopted the same convention as for the $\overline{\bf 5}$'s, namely that
$Q_{\rm horiz}=\frac{1}{2}$ goes with the $SU(2)_L$ doublet fermions and singlet bosons, and vice-versa for $Q_{\rm horiz}=0$. 
Note that anomaly cancellation (such as cancellation of  the $SU(3)^2 U(1)_{\rm horiz}$ anomalies) 
requires the existence of a {\it second} $e$-multiplet 
which we do not show with the horizontal charges negated. 
This will be present in the explicit example to be presented in Sect.~\ref{section4}.

For vector-like pairs, in particular the Higgses, anomaly cancellation is achieved 
with a slightly different entwining in which the supersymmetry transformation is coupled with a permutation of $SU(2)_L$ with $SU(3)$ fundamental.
Let us assume that there is a vector-like pair of Higgs supermultiplets ${ {\bf 5}}_{h_u,\frac{1}{2}}$ and ${ \overline{\bf 5}}_{{h}_d,-\frac{1}{2}}$. In contrast with the matter ${\bf 5}$'plets, entwined SUSY leaves light the scalar doublets  $h_u$ and ${h}_d$ as well as a vector-like pair of fermionic color triplets. The entwined multiplet takes the form 
\begin{equation}
 \label{5higgs}
{\overline{\bf 5}}_{h_d,e}\oplus {\overline{\bf 5}}_{h_u,e} \,\, =\,\,
\left\{
\begin{array}{rrl}
{\mbox {fermions}}: & ({\overline{{\bf 3}}} ,1)_{-\frac{1}{2}}\oplus ({{{\bf 3}}} ,1)_{\frac{1}{2}} ~&\equiv ~ \tilde{T}_{d,-\frac{1}{2}}    \oplus  \tilde{T}_{u,\frac{1}{2}}  \\
{\mbox {bosons}}: &  ({{\overline{\bf 2}}} ,1)_{-\frac{1}{2}}\oplus (1, {{\bf 2}})_{\frac{1}{2}} ~&\equiv ~ {h}_{d,-\frac{1}{2}} \oplus  h_{u,\frac{1}{2}}  ~ ,
\end{array}\right.
\end{equation}
where $T$ and $\tilde{T}$ are color triplets.
Of course the $e$-SUSY structure
is a feature of the matter and Higgs sectors only.  In particular, it does
not extend to the gauge sector;
indeed, the gauginos are heavy, as we have said, and there are no partners for the gauge bosons. 

Entwined SUSY is not just a property of the spectrum --- it also governs the allowed couplings.
As an example, let us consider the part of the superpotential of the original GUT theory that 
encapsulates the down and lepton Yukawa couplings:
\begin{equation} 
\label{eq:d-yuk}
W_{\rm d-yuk} ~=~ \sqrt{2} \, g_{4} \left ( {{\bf 10}}_{\frac{1}{2}} \,{ \overline{\bf 5}}_{0}\, { \overline{\bf 5}}_{h_d,-\frac{1}{2}} +  { {\bf 10}}_{0} \,{ \overline{\bf 5}}_{\frac{1}{2}} \,{ \overline{\bf 5}}_{h_d,-\frac{1}{2}}\right) \, .
\end{equation} 
Note that both terms share the same coupling $g_4$.
We have also assumed
that the Higgs descends from a higher-dimensional gauge boson and thus is an off-diagonal component of the adjoint representation of a larger, broken symmetry.   Indeed, this is always the case if the Higgs 
is a state from the Neveu-Schwarz Neveu-Schwarz (NS-NS) sector;  as we shall shortly see,
such states are generic.

We can divide the superpotential 
in Eq.~(\ref{eq:d-yuk})
into two components, $W_{\rm d-yuk} = W_{f}+W_{b}$, where 
$W_{f}$ involves those matter supermultiplets whose fermions remain light after SUSY breaking 
while $W_{b}$ involves the supermultiplets whose bosons remain light.  
Keeping only those pieces that include the Higgs doublets, 
we then have
\begin{eqnarray} 
W_f & =& \sqrt{2} \, g_{4} \left ( q_{\frac{1}{2} } \,d^c_{0 }\, h_d +  e^c_{0 }\, \ell_{\frac{1}{2} }\, h_d  \right) \nonumber \\
W_b & =& \sqrt{2} \, g_{4} \left (  e^c_{\frac{1}{2} } \, \ell_{0 } \, h_d+ q_{0 } \,d^c_{\frac{1}{2} }\,h_d   \right) \, ,
\end{eqnarray} 
where $q,\ell,d^c,e^c,h_d$ denote the complete supermultiplets. 
However, the crucial point is that the scalar Higgs $h_d$ 
in the light theory couples to an entire $e$-multiplet, {\it i.e.}\/, 
both to fermions and to their $e$-partners, with degenerate couplings. 
Thus, for example, the quadratically divergent contributions to the Higgs squared-mass from these multiplets
still cancels, 
much as in genuine SUSY.~ 
Indeed, this $e$-SUSY structure occurs for every pair of couplings in the  
original GUT theory that are already {\it independently}\/ invariant under a permutation of $Q_{\rm horiz}$.  
For example, if only the first piece of the superpotential in Eq.~(\ref{eq:d-yuk}) had existed, 
then only the first pieces of $W_b$ and $W_f$ would have been present, and $e$-SUSY would have been broken in these couplings. 
A typical theory contains examples of both kinds of coupling.  
In this connection, we remark that such a structure
also has the potential to solve the 
Higgs hierarchy problem.  We shall comment on this below.

At first sight, the emergence of {\it e}\/-SUSY may seem surprising. 
However, in the present context,
this structure is essentially forced upon us.
To see why, we begin 
by recalling that in our scenario, we are compactifying from ten dimensions to four dimensions
in two stages:   first $10\to 4+\delta$, and subsequently $4+\delta\to 4$.
Moreover, the existence of the GUT-precursor structure means that 
only the second stage of the compactification may break the
GUT symmetry.  Indeed, if this were not the case, then there would be ${\cal O}(M_s)$ mass splittings 
between components of 
a single GUT multiplet. 
However, there are two components to this final stage of compactification.
The first is an action ({\it i.e.}\/, a set of phases) on the fields associated with the 
Scherk-Schwarz twist.    By itself, this would break the 4D theory from ${\cal N}=2$~SUSY
to ${\cal N}=0$~SUSY, resulting in a non-chiral theory.    By contrast, the second ingredient is the aforementioned
orbifold action which, by itself, would break ${\cal N}=2$~SUSY to ${\cal N}=1$~SUSY.~    In principle, either of these
is a suitable place in which to embed the breaking of the GUT symmetry, and indeed
both options lead to ${\cal O}(M_{\rm KK})$ mass splittings between different components of a single GUT multiplet.
In this paper, we will without loss of generality assume that the GUT symmetry is broken via the latter procedure.

Let us now consider the properties of the resulting spectrum.
Relative to the orbifold action of $\mathbb{O}_d$,
some sectors of the theory will be untwisted and some will be twisted.
Of course, the twisted sectors remain fully supersymmetric because they are blind to the large radius.
Thus we immediately see that it is only the untwisted sectors which exhibit the SUSY-breaking
and the eventual entwined SUSY.~
As discussed above, the GUT-precursor structure indicates that any SUSY- and GUT-breaking that occurs in these 
sectors must be driven entirely by non-trivial GSO phases --- {\it i.e.}\/, by a non-trivial Scherk-Schwarz action. 
The effect of these phases is to  lift the masses of certain components of the 
SUSY GUT multiplets to ${\cal O}(1/R)$, where $R$ is the typical compactification radius. 

We can gain a simple understanding of which components will have their masses lifted as follows.
In general, each string state has a corresponding charge vector which we may write in the form
\beq
\label{appcharges}
{\bf Q}  ~=~  ( Q_{\rm s.t.} , Q_{R} ~ | ~  Q_{U(3)}, Q_{U(2)}, {\bf Q}_{\rm horiz} )\, .
\eeq
Here $Q_{U(3)}$ and $Q_{U(2)}$ denote the charges of this state under the Standard-Model $U(3)$ and 
$U(2)$ gauge symmetries, while ${\bf Q}_{\rm horiz}$ generally denotes charges under
other gauge symmetries which may be viewed as horizontal relative to those of the Standard Model.
Likewise,
$Q_{\rm s.t.}$ are charges indicating the spacetime helicity (spin-statistics) of the state,
while $Q_R$ denote its internal $R$-charges.

Note that these charge vectors have natural identifications within the special case of the heterotic string.
For heterotic strings in six dimensions, the charge vectors of the string states
fill out a $(8,20)$-dimensional Lorentz self-dual lattice
where the 20 dimensions correspond to the bosonic (or gauge) side of the heterotic string
and the 8 dimensions correspond to the superstring (or spacetime) side.
For SM-like strings, we may in general identify 
three dimensions within the 20  
as corresponding to $U(3)$ charges,
while we may identify two others as corresponding to $U(2)$.
Relative to these gauge groups, we may regard the remaining 15 dimensions
as corresponding to horizontal symmetries.
These three sets of charges therefore correspond 
to $Q_{U(3)}$, $Q_{U(2)}$, and ${\bf Q}_{\rm horiz}$.
Likewise, on the spacetime side, 
two lattice dimensions correspond to the spacetime helicities (spin-statistics).
These are therefore the charges we have denoted ${Q}_{\rm s.t.}$.
Note that the remaining six dimensions on this side 
are purely internal, and their charges may be viewed as $R$-charges.    
String consistency constraints concerning the worldsheet supercurrent 
correlate these charges with ${Q}_{\rm s.t.}$,
and thus the $R$-charges, like $Q_{\rm s.t.}$, are sensitive to whether a given
state is bosonic or fermionic in spacetime.

In general, different states within a given GUT multiplet will have different charge vectors.
For example, a complete supermultiplet $\chi$ in the GUT theory might decompose into fermionic and bosonic 
pieces with charge vectors
\begin{eqnarray}
\label{eq:qhoriz}
{\mathbf Q}_\chi &=& ( { \scriptstyle\frac{1}{2}} ,{ \scriptstyle\frac{1}{2}} ~ | ~  Q_{U(3)}, Q_{U(2)}, {\bf Q}_{\rm horiz}) \nonumber \\
{\mathbf Q}_{\tilde \chi} &=& ( 0 , 0 ~ | ~  Q_{U(3)}, Q_{U(2)},   {\bf Q}_{\rm horiz} ) \, ,
\end{eqnarray}
where we have left $Q_{U(3)}, Q_{U(2)}$ unspecified in order to allow for different Standard-Model charges. 
In an untwisted sector the bosons typically have
$Q_{\rm s.t.} = Q_{R} =0$, which we have adopted above for concreteness.

The question is therefore to determine which of the
massless states in such multiplets will have their masses lifted 
by the Scherk-Scherk twist.
In general, for a given field $\psi$ and a compactified direction with coordinate $x_5$ and radius $R$, 
such a twist takes the form
\beq
   \psi(x_5+2\pi R)~=~ e^{i\pi {\bf e}\cdot {\bf Q}} \,  \psi(x_5)~,
\label{twistt}
\eeq
where the vector ${\bf e}$ specifies the particular twist
and takes the form
\begin{eqnarray}
{\bf e}  &=& ( e_{\rm s.t.} , e_{R} ~ | ~  e_{U(3)}, e_{U(2)}, {\bf e}_{\rm horiz} )\, ,
\end{eqnarray}
and where the dot-product is Lorentzian ({\it i.e.}\/, gauge minus spacetime).
In general, implementing this twist raises the masses of those states which carry a net Scherk-Schwarz charge.
More specifically, 
in the presence of a universal compactification radius $R$,  
we see from Eq.~(\ref{twistt}) that
any state with a charge vector ${\bf Q}$ will experience a shift in its 
KK mode number $k\in \IZ$ of the form $k \to k + {\bf e}\cdot {\bf Q}$.
This implies that the mass of a previously massless state with charge vector ${\bf Q}$ now becomes 
\begin{equation}
    m ~=~ \frac{|\Delta k|}{R} 
    ~~~~{\rm  where}~~  \Delta k ~\equiv~ {\bf e\cdot Q}~~ {\rm mod}~(1)~.
\label{masshift}
\end{equation}
Of course, if ${\bf e}\cdot {\bf Q}$ is an integer, then the KK mode numbers
of our states merely shift by an integer.  There is thus always another KK mode 
$k'= -{\bf e}\cdot {\bf Q}\in \IZ$ which now becomes massless and which takes the place of the
original state in the sense that it has the same spacetime properties.
This then explains the restriction ``mod (1)'' in Eq.~(\ref{masshift}). 
We conclude that only those states for which ${\bf e}\cdot {\bf Q}\in \IZ$
survive in the massless spectrum.

Given this, each choice of ${\bf e}$-vector corresponds
a specific resulting pattern of SUSY-breaking and GUT-breaking.
Certain aspects of the required twist are then obvious.
First, the Scherk-Schwarz twist has to distinguish bosonic states from fermionic states.
In principle, this can be accomplished by having this twist be sensitive to
either ${Q}_{\rm s.t.}$ or ${ Q}_R$;   for technical reasons the choice ${Q}_R$ is more 
natural.
Likewise, in order to break the GUT group, the twist must be sensitive to ${Q}_{U(3)}$
or ${Q}_{U(2)}$.
Finally, string self-consistency conditions then require that the twist also generically act on the different horizontal charges of the states ${\bf Q}_{\rm horiz}$.

As an example,  let us suppose that we wish the Scherk-Schwarz twist to act on $Q_{R}$ (in order to distinguish
bosons from fermions), and to simultaneously break the GUT symmetry by acting on $Q_{U(2)}$
and ${\mathbf Q}_{\rm horiz}$ but not on $Q_{U(3)}$. A relevant Scherk-Schwarz action for the projection above is then given by the vector 
\begin{eqnarray}
{\bf e}  &=& ( 0~ , 1 ~ | ~  0~, 1 ~, {\bf e}_{\rm horiz} )\, .
\end{eqnarray}
The states from our original supermultiplet that remain light therefore obey   
\begin{eqnarray}
\label{eq:qrule}
 Q_{U(2)}+Q_{\rm horiz}  - Q_R   ~=~ 0\,\,\, ~~~\mbox{mod (1)}\, ,
\end{eqnarray}
where we define $Q_{\rm horiz}\equiv {\bf e}_{\rm horiz}\cdot {\bf Q}_{\rm horiz}$. 
Clearly there is some model-building freedom in the choice of ${\bf e}_{\rm horiz}$  and the corresponding distribution of $Q_{\rm horiz}$ charges among the matter multiplets.

In order to determine the resulting massless spectrum, 
it is necessary to discuss the values of the trace $U(2)$ charge $Q_{U(2)}$, which depends on how the representations are constructed. In a typical construction the matter ${\bf 10} $ and $\overline{\bf 5}$ come from spinor representations of a larger gauge group [{\it e.g.}\/,  $SO(16)$], and in this case  the matter doublets $q$ and  $\ell$ have $Q_{U(2)}= 0$ while $e^c$ and $d^c$ have $Q_{U(2)}= \frac{1}{2}$ and $u^c$ and $\nu^c$ have  $Q_{U(2)}= -\frac{1}{2}$. Meanwhile, given our previous  assumptions,  the Higgses appear in the bifundamental (with one factor in the $SU(2)_L$ group and the other factor in a hidden gauge group).  They therefore have $Q_{U(2)}=\pm\frac{1}{2}$.

Adopting these charges and applying the projection in Eqs.~(\ref{eq:qrule})  with the charges in Eq.~(\ref{eq:qhoriz}), 
we see that the remaining light matter fermions are  $q_{\frac{1}{2}},~\ell_{\frac{1}{2}},~u^c_{0},~d^c_{0},~  \nu^c_{0},~  e^c_{0}$ (and Higgs triplets).  Likewise, the light scalars are 
$h_d,~h_u,~\tilde{q}_{0},~\tilde{\ell}_{0},~\tilde{u}^c_{\frac{1}{2}},~\tilde{d}^c_{\frac{1}{2}},~  \tilde{\nu}^c_{\frac{1}{2}},~  \tilde{e}^c_{\frac{1}{2}}$.  Thus, the massless left-handed fermion matter doublets have $Q_{\rm horiz}=\pm \frac{1}{2}$ after applying the Scherk-Schwarz mechanism, while their bosonic ``pseudo-superpartners'' actually come from the states with $Q_{\rm horiz}=0$. 
Conversely the massless right-handed matter fermions have $Q_{\rm horiz}=0$  and their bosonic ``pseudo-superpartners'' have $Q_{\rm horiz}=\frac{1}{2}$.   All other states acquire a mass $1/2R$. 
Finally,  the original GUT theory must be free of $SU(5)^2\times U(1)_{\rm horiz}$ anomalies, which requires that
two $e$-twisted generations descend from four GUT generations 
(with corresponding horizontal charges $0,\, \frac{1}{2},\,0,-\frac{1}{2}$). 

This is precisely the {\it e}\/-SUSY structure described above. 
The specific distribution of charges may differ from theory to theory, 
but the general robust feature is that differently-charged components comprise $e$-supermultiplets.
Indeed, the particular distribution of charges described above is quite typical. 
The crucial feature of the Scherk-Schwarz mechanism that results in this structure is that the breaking of supersymmetry and gauge group occur simultaneously in the underlying string construction.  This leads to a correlation between $Q_{\rm horiz}$ and the $R$-charges, and hence between $Q_{\rm horiz}$ and the spacetime spins of the components of the $e$-supermultiplets that actually descend from different supermultiplets of the original 
SUSY GUT. 

Note that this mechanism operates only for supermultiplets that carry horizontal charges overlapping with the Scherk-Schwarz action ${\bf e}$.~  States such as gauginos and gravitinos
that do not carry horizontal $U(1)$ charges are simply made heavy, whereas the untwisted chiral matter supermultiplets are typically projected in half into such entwined-supermultiplets by the Scherk-Schwarz twist in the manner 
described above.
It is also important to note that at large volumes, the horizontal $U(1)$ symmetries will be exceedingly weak:   
their gauge couplings (like all of the gauge couplings in such large-volume compactifications) 
begin extremely small at the string scale, but because $U(1)$ symmetries have $\tilde b>0$, 
their couplings become even smaller upon running down to lower energies.
Thus, for all practical purposes, such $U(1)$ symmetries can be treated as effectively global 
at low energies. 

A final remark is in order.
Given the above discussion, it is clear that left- and right-handed SM fields are projected differently. 
As we shall see in detail through our consideration of an explicit model, 
in order to maintain anomaly cancellation one must therefore be careful to include sectors that are twisted under the orbifold but also have a Scherk-Schwarz twist. 
The states arising from such sectors are typically SM singlets and are 
evident in the effective 6D theory at small radius, where the vector ${\bf e}$ has the same status as the other projection vectors in the theory.  This has been discussed recently in Ref.~\cite{Aaronson:2016kjm}.~  Due to their orbifold-twisted nature, these states cannot gain a mass from the interpolation to large radius. 
They therefore fall into genuine supermultiplets and do not contribute to the one-loop cosmological constant.  
Thus, these sectors do not change our conclusion,
stated above, that the entwined-SUSY
structure is a feature of the untwisted sectors alone.

\section{Constructing a metastable non-SUSY heterotic Standard Model}
\label{section4}

We now present a non-supersymmetric  heterotic string model that displays the features described above 
while at the same time exhibiting the required boson/fermion degeneracy of the massless modes,
as required for tadpole stability in our construction. 
Our primary purpose here is to explicitly demonstrate that the features claimed in the 
previous sections indeed occur within the context of a fully self-consistent metastable non-supersymmetric
heterotic string model.
Moreover, this model is also phenomenologically semi-realistic:  
it incorporates the fundamental
building blocks of the Standard Model such as the Standard-Model gauge group, complete chiral
generations, 
and Higgs fields, all without supersymmetry. 
The model does not give rise to massless gravitinos and gauginos, and thus SUSY is indeed broken.
This model nevertheless displays both the GUT-precursor
structure and the {\it e}\/-SUSY structure, all while retaining one-loop stability 
with a near-vanishing one-loop cosmological
constant.  
This model thus provides a concrete and illustrative example of the ideas presented
in Sect.~\ref{entwined}, even if it is not fully realistic at a phenomenological level.
This model will also provide us with a template for discussing 
some general phenomenological properties of {\it e}\/-SUSY, such as the
existence of an approximate moduli space and the prevention of rapid proton decay. 
These issues will be discussed in Sect.~\ref{pheno}.~
However the precise construction in this 
section will not be crucial for understanding the more general points
in Sect.~\ref{pheno},
and thus the reader unconcerned with the specifics of the model to be presented in this section
can proceed directly to Sect.~\ref{pheno}.

\subsection{Building a suitable string model:   Basic architectural approach \label{subsectA}} 

We begin the construction of our model along the lines described in previous sections.
Specifically, following Ref.~\cite{Abel:2015oxa}, we generate the compactification manifold $\mathbb{K}$ within 
the free-fermionic formalism~\cite{FF1,FF2,FF3}, while $\mathbb{O}_\delta$ is taken to be
$\mathbb{T}_2/\mathbb{Z}_2$ (again noting that the $\mathbb{Z}_2$ acts also on the  $\mathbb{K}$). 
The 6D model is defined by a set of basis vectors $V_{i}$ 
which describe the phases of worldsheet fermions in each sector
and which collectively
generate the charges schematically indicated in
Eq.~(\ref{appcharges}).  
In particular, these basis vectors $V_i$ have rank 8 on the spacetime side and 20 on the internal gauge side,
and give rise to a spectrum of states whose 28-dimensional charge vectors ${\bf Q}$ 
were discussed in Sect.~\ref{entwined}.
The model is also specified through a set of so-called ``structure constants'' $k_{ij}$ (or GSO phases) which collectively determine the particular chiralities and redundancies in the model. 
The orbifolding compactification to 4D is then accompanied by an action (set of phases) 
on the fermions described by a vector of charges $b_3$.  
The central ingredient in the compactification is the introduction of the Scherk-Schwarz action following the so-called Coordinate-Dependent Compactification (CDC) method of Refs.~\cite{Ferrara:1987es,Ferrara:1987qp,Ferrara:1988jx,Kounnas:1989dk, Antoniadis:1992fh}. This spontaneously breaks supersymmetry  but retains the desirable properties of the original theory, in particular modular invariance, misaligned SUSY, and hence one-loop finiteness.  

As outlined in Sect.~\ref{entwined},  the CDC is described by a vector $\mathbf{e}$ which encompasses a {\it discrete}\/ $U(1)$ rotation of the Lorentzian compactification lattice that depends on the $\mathbb{T}_{2}$ coordinates and leaves the worldsheet supercurrent invariant. However the {\it local}\/ generator of this rotation does not commute with the worldsheet supercurrent [{\it i.e.}\/, it lives partly in the $SO(4)$
subgroup associated with the compactification from 10D to 6D so it involves the $R$-symmetry], and hence supersymmetry is spontaneously broken. As a consequence  the graviton remains massless but the gravitinos pick up a mass of order $1/2R$ where $R$ is the
generic compactification scale. 
The CDC may simultaneously act upon the internal 
worldsheet fermions, breaking gauge symmetries, and in the present context this is the source of GUT breaking. The broken gauge bosons as well as the gauginos of the unbroken gauge groups gain a mass while their superpartners remain massless. 

As far as the massless spectrum is concerned, 
${\bf e}$ can be treated as just another set of projection phases. 
Indeed in the $R\rightarrow0$ limit, the role of $\mathbf{e}$ explicitly reverts to that of the other $V_{i}$, as discussed in detail in 
Ref.~\cite{Aaronson:2016kjm}. As we shall see later, this allows a simple determination of the additional states 
which are present --- states  which are required for anomaly cancellation and which come from sectors twisted under both the CDC and the orbifold. 

One feature that makes finding a suitable model feasible is that the orbifold and the CDC act somewhat independently of each other. Only the untwisted orbifold sectors depend on the radius
and therefore feel the CDC, while the twisted sectors remain supersymmetric (until they too pick up one-loop radiative corrections via what is essentially gauge- or gravity-mediation). Conversely the
orbifold acts merely to remove half the states of the untwisted theory, such that if, {\it e.g.}\/, one finds a non-orbifolded theory that exhibits a boson/fermion degeneracy, then up to a choice of $k_{ij}$ it follows that 
there exists an {\it orbifolded}\/ theory that also exhibits the boson/fermion degeneracy but with half the untwisted particle content. This particular issue is explained in more detail in
Appendix~\ref{appendix: 1}.

Despite these simplifications, the multiple consistency conditions 
add up to a relatively constraining set of requirements for any consistent model.  
Collected together, these constraints are as follows:
\begin{itemize}
\item The basis vectors $V_i$ and structure constants $k_{ij}$ must satisfy the usual modular-invariance constraints for the original 6D theory~\cite{FF1,FF2,FF3}.
\item  The $\mathcal{N}=0$ model must interpolate between the $\mathcal{N}=1$ 6D model presented above in the $R\rightarrow \infty$ limit and a different 6D model in the $R\rightarrow 0$ limit (which, in order to break the GUT symmetry,
must also be SUSY-breaking~\cite{Aaronson:2016kjm}). 
In the $R\rightarrow 0$ limit, the CDC vector $\mathbf{e}$ reverts to the role of a normal vector and is added 
in the spin structure of that model~\cite{Aaronson:2016kjm}.   
Therefore, the $\mathbf{e}$ vector must obey the {\it same}\/ modular-invariance constraints as the other $V_i$ vectors that define the model in 4D.~ A simple way to do this is to impose the constraint $\mathbf{e}\cdot\mathbf{e}=1~\mbox{mod}(2)$, where the conventions for inner products are as presented in
Ref.~\cite{Abel:2015oxa} and where $k_{ie}=0$ \cite{Aaronson:2016kjm}.

\item Removing the CDC must restore supersymmetry. In other words, supersymmetry breaking should be the result of a mismatch between an $\mathcal{N}=1$ supersymmetry
left invariant by the orbifold and a {\it different}\/ $\mathcal{N}=1$ supersymmetry left invariant by the CDC.~ This mismatch is governed by the choice of structure constants $k_{ij}$.

\item There must exist an alternative choice of structure constants $k_{ij}$ that can also restore supersymmetry.

\item In at least one twisted sector of the orbifold, there must exist a basis in which the orbifold acts as a conjugation on the charge lattice, which in conjunction with the 
fact that the orbifold reverses all KK and winding numbers, results in a consistent projection on the spectrum. 
Note that overlaps of complex phases with the CDC vector require special treatment (see
Appendix~\ref{appendix: 1}).
\end{itemize}

In addition to the above, we may now also impose two further constraints:
\begin{itemize}
\item In order to realize the GUT-precursor structure, only the CDC may break the GUT symmetry.

\item The resulting model must exhibit a boson/fermion degeneracy for all massless modes.
\end{itemize}

Finding working solutions that satisfy all of these constraints is highly non-trivial. 
It is therefore remarkable that one can find a theory which not only satisfies all of these constraints but
which also resembles the Standard Model. 
In fact  we can make very general statements 
about the form of such a model, and about where the matter generations and the Higgses will appear. 
In order to do this we need to extend the discussion of Ref.~\cite{Abel:2015oxa} to include cases where both the CDC vector 
$\mathbf{e}$ and the orbifold vector $b_3$ overlap more general complex GSO phases of the compactification to 6D.~ 
It is less obvious how to define a consistent basis in these cases, 
and this particular issue is also treated in  Appendix~\ref{appendix: 1}. The upshot of that discussion is that in order to determine the physical spectrum one can perform projections as if the
vectors were all defineable in the same complex basis.  This is ultimately because the interaction of the orbifold with the CDC is somewhat trivial: the orbifold simply projects out half the untwisted states as above,
while twisted sectors do not feel the CDC.  The minimal consistent unit that emerges is a pair of complex worldsheet fermions with phases
\begin{equation}
\label{toot}
 V_n\supset -\left(\frac{1}{4},\frac{1}{4}\right)~,~~~~~~~
      b_{3}  \supset  -\left(0,\frac{1}{2}\right)~,~~~~~~~ 
      \mathbf{e}  \supset  \left(\frac{1}{2},\frac{1}{2}\right)~.
\end{equation}

With this in mind, we can sketch how the vectors for an SM are expected to divide into their crucial blocks. With only a minor loss of generality, the structure one requires for the phases (in units of $2\pi$) is given by
\begin{table}[H]
\centering
\begin{tabular}{clccccr}
Group = & $ [ $ \small{ $ \psi_{\rm s.t.}^1 \psi_{\rm s.t.}^{2} $} & $...|...$ & {\small{ $U(3)\times U(2) $}} & {\small $ U(1)^3$} & $... ] $\nonumber \\
\hline 
$V_0$  & $ [ $   $~~~ 1~~~1 $ & $...|...$ &$ 1~1~1~1~1 $ & $ 1~1~1$ & $... ]  $ \\
$V_1$  & $ [ $   $~~~ 0~~~0 $ & $...|...$ &$ 1~1~1~1~1 $ & $ 1~1~1$ & $... ]  $ \\
$V_{i=2,...,n-1}$  & $ [ $   $~~~ 0~~~0 $ & $...|...$ &$ 0~0~0~0~0 $ & $ *\,*\,*$ & $... ]  $ \\
$V_n$  & $ [ $   $~~~ 0~~~0 $ & $...|...$ &$ \mbox{\small{$\frac{1}{2}$\hspace{1mm}}} \mbox{\small{$\frac{1}{2}$\hspace{1mm}}} \mbox{\small{$\frac{1}{2}$\hspace{1mm}}}
\mbox{\small{$\frac{1}{2}$\hspace{1mm}}} \mbox{\small{$\frac{1}{2}$\hspace{1mm}}}  $ & $ \mbox{\,\small{$\frac{1}{2}$\hspace{1mm}}} \mbox{\small{$\frac{1}{2}$\hspace{1mm}}}
\mbox{\small{$\frac{1}{2}$\hspace{1mm}}} $ & $... ]  $ \\
$b_{3}$  & $[ $   $~~~ 1~~~0 $ & $...|...$ &$ 1~1~1~1~1 $ & $ 0~1~0$ & $... ]  $ \\
{\bf e}  & $[ $   $~~~ 0~~~0 $ & $...|...$ &$ 0~0~0~1~1 $ & $ 1~0~1$ & $... ]  $ \\
\end{tabular}
\end{table}
\noindent Here there are 8 entries on the spacetime side and 20 on the gauge side, and 
all entries are to be understood as multiplied by $-1/2$.
Comparing with the notation
in Eq.~(\ref{appcharges}),
we see that ${Q}_{\rm s.t.}$ corresponds to $\psi^{1,2}_{\rm s.t.}$.
Note that the notation `$*$' denotes a wildcard $0$ or $1$.
The boundary-condition phases of the left-moving (gauge-side) fermions 
in the basis vectors $V_{i=2..n-1}$ break the gauge symmetry at the string scale 
to $SU(5)\times U(1)_1 \times U(1)_2 \times U(1)_3 \times \ldots$, because they act 
degenerately on the components of the GUT.~ 
The components denoted `$*$' allow these vectors to break the $SO(6)$ gauge factor at the
string scale. The vector $V_{n}$ is required to break the gauge symmetry to unitary groups, with in this case 
$-1/4$ complex phases on the worldsheet fermions. 
This, then, is the 6D supersymmetric GUT.~ 
The vector $b_3$ which accompanies the orbifolding projects the theory down to ${\cal N}=1$ in 4D.~ 
In the absence of the Scherk-Schwarz action, this gives a ``parent'' GUT model from which one may deduce the embedding of the multiplets of the eventual SM-like model.  This is a useful check. 

The overlap of ${\bf e}$ is then prescribed by Eq.~(\ref{toot}).
Choosing ${\bf e}$ to give the final breaking of the $SU(5)$ symmetry and to break the spacetime SUSY 
allows us to retain untwisted Higgs states which appear as a bi-fundamentals.
These states are essentially components of the higher-dimensional gauge fields.  
Such fields naturally have phenomenologically attractive shift symmetries (as evident in the effective tree-level theory of Ref.~\cite{Abel:2016hgy} but  broken at one-loop level~\cite{Abel:2015rkm}). 
These untwisted Higgses will be found in the NS-NS sector, where they will generically have the form 
\begin{eqnarray}
h_d & ~\sim~ & \psi_{\rm s.t.}^2 |0\rangle\otimes\tilde{\psi}_{U(2)}\tilde{\psi}_{U(1)_{1,3}}^{\dagger}|0\rangle~\nonumber \\
{h}_u & ~\sim~ & \psi_{\rm s.t.}^2 |0\rangle\otimes\tilde{\psi}_{U(2)}^{\dagger}\tilde{\psi}_{U(1)_{1,3}}|0\rangle~ .
\label{Higgsesreps}
\end{eqnarray}
Note that in the notation of Ref.~\cite{FF3} we are using $\psi \supset {b_{n>0} \oplus d_{n>0}^\dagger }$ 
and $\psi^\dagger  \supset {b^\dagger_{n>0} \oplus d_{n>0} }$. The projections in this
sector are very general.  That from the orbifold takes the form $b_{3}\cdot N_{\mathbf{0}}={1}/{2}$, and hence there are no Higgs states involving $\tilde{\psi}_{U(1)_{2}}$. 
However there are light vector-like triplets of the form 
\begin{equation} 
T~\sim~\psi_{\rm s.t.}^2 |0\rangle\otimes\tilde{\psi}_{U(3)}\tilde{\psi}_{U(1)_{2}}^{\dagger}|0\rangle ~
\label{Higgsreps2}
\end{equation}
plus its conjugate.  Note this vector-like pair of Higgs {\bf 5}-plets is precisely in the form of Eq.~(\ref{5higgs}).
Another benefit of choosing $\mathbf{e}$ to break the GUT symmetry is now evident: states with $\mathbf{e\cdot Q}=1/2$ mod$\,$(1) are lifted by the CDC, so if ${\bf e}$ were degenerate 
across the $U(3)\times U(2)$ entries, then it would either lift the Higgs mass or leave a larger $SU(4)$ symmetry.

\subsection{A full non-supersymmetric metastable SM-like 4D string model}

Thus far we have only presented the basic architectural ``kernel'' of a successful model.
However, given this, we can now proceed to construct 
a full string model that meets all of the requirements listed above
while also satisfying all of the constraints that we would expect
for a full, self-consistent string model.
To do this, we shall begin by presenting the parent 4D ${\cal N}=1$ GUT-like model 
from which our eventual SM-like model descends, as described above.
We shall then present our final, SM-like model.

\begin{table}[H]
\centering
\resizebox{\textwidth}{!}{
\begin{tabular}{||c||*{1}c|{l}||}
\hline
\hline
~Sector~& {\small $\xx \psi^{34}\yyr \psi^{56}\yyr \chi^{34}\:y^{34}\yyr\omega^{34}\yyr\chi^{56}y^{56}\yyr\omega^{56}$} &{\small $\overline{y}^{34}\yyr\overline{\omega}^{34}
\yyr\overline{y}^{56}\yyr\overline{\omega}^{56}\yyr\overline{\psi}^{1}\yy\overline{\psi}^{2}\yy\overline{\psi}^{3}\yy\overline{\psi}^{4}\yy\overline{\psi}^{5}\yy
\overline{\eta}^{1}\yy\overline{\eta}^{2}\yy\overline{\eta}^{3}\yy\overline{\phi}^{1} \yy\overline{\phi}^{2}\yy\overline{\phi}^{3}\yy\overline{\phi}^{4}\yy\overline{\phi}^{5}
\yy\overline{\phi}^{6}\yy\overline{\phi}^{7}\yy\overline{\phi}^{8}$} \\ 
\hline
\hline
$V_0$&1\xx 1\xx 1\xx 1\xx 1\xx 1\xx 1\xx 1\xx &1\xxl  1\xxl  1\xxl 1\xxl  1\xxl 1\xxl 1\xxl  1\xxl 1\xxl 1\xxl 1\xxl 1\xxl 1\xxl 1\xxl 1\xxl 1\xxl 1\xxl 1\xxl 1\xxl 1 \\
$V_1$&0\xx 0\xx 0\xx 1\xx 1\xx 0\xx 1\xx 1\xx &1\xxl  1\xxl  1\xxl 1\xxl  1\xxl 1\xxl 1\xxl  1\xxl 1\xxl 1\xxl 1\xxl 1\xxl 1\xxl 1\xxl 1\xxl 1\xxl 1\xxl 1\xxl 1\xxl 1 \\
$V_2$&0\xx 0\xx 1\xx 0\xx 1\xx 1\xx 0\xx 1\xx &0\xxl  1\xxl  0\xxl 1\xxl  0\xxl 0\xxl 0\xxl  0\xxl 0\xxl 0\xxl 1\xxl 1\xxl 1\xxl 1\xxl 1\xxl 1\xxl 1\xxl 1\xxl 1\xxl 1 \\
$b_3$&1\xx 0\xx 1\xx 0\xx 0\xx 0\xx 0\xx 1\xx &0\xxl  0\xxl  0\xxl 1\xxl  1\xxl 1\xxl 1\xxl  1\xxl 1\xxl 0\xxl 1\xxl0\xxl 1\xxl 0\xxl 1\xxl 0\xxl 0\xxl 0\xxl 1\xxl 1 \\
$V_5$&0\xx 0\xx 0\xx 0\xx 0\xx 0\xx 1\xx 1\xx &1\xxl  0\xxl  0\xxl  1\xxl  0\xxl  0\xxl 0\xxl  0\xxl 0\xxl 0\xxl 1\xxl1\xxl 0\xxl 1\xxl 1\xxl 1\xxl 1\xxl 0\xxl 1\xxl 1 \\
$V_7$&0\xx 0\xx 0\xx 1\xx 1\xx 0\xx 0\xx 0\xx &0\xxl  1\xxl  0\xxl  1\xxl  $\frac{1}{2}$\xxh  $\frac{1}{2}$\xxh $\frac{1}{2}$\xxh  $\frac{1}{2}$\xxh $\frac{1}{2}$\xxh $\frac{1}{2}$\xxh $\frac{1}{2}$\xxh$\frac{1}{2}$\xxh 1\xxl 1\xxl 1\xxl 1\xxl 1\xxl 0\xxl 1\xxl 0 \vspace{0.8pt} \\
\hline
\hline
\end{tabular}}
\caption{~Spin structure of the parent ~$\mathcal{N}=1$ ~4D GUT model. 
This spin structure is accompanied by two bosonic degrees of freedom compactified 
on a ~$\mathbb{Z}_2$ orbifold with twist action corresponding to the vector ~$b_3$.
For notational simplicity, each entry in this table is to be understood as multiplied by $-1/2$.}
\label{table:A}
\end{table}

As discussed above, our parent 4D ${\cal N}=1$ GUT-like model can be specified through those quantities that
uniquely generate it:   a set of boundary-condition vectors as well as a corresponding matrix $k_{ij}$
of projection phases.
In the present case, the boundary-condition vectors are 
given in Table~\ref{table:A}, while the $k_{ij}$ projection phases
(``structure constants'') are given in the $\lbrace V_0,V_1,V_2,b_3,V_5, V_7\rbrace$ basis by
\begin{equation}
 k_{ij}~=~
\begin{pmatrix}
 0 & 0 & 0 & 0 & 0 & 0  \\ \vspace{4pt}
 0 & 0 & 0 & 0 & 0 & 0  \\ \vspace{4pt}
 0 & \frac{1}{2} & 0 & \frac{1}{2} & 0 & \frac{1}{2} \\ \vspace{4pt}
 0 & \frac{1}{2} & \frac{1}{2} & \frac{1}{2} & 0 & \frac{3}{4} \\ \vspace{4pt}
 0 & 0 & 0 & 0 & 0 & \frac{3}{4} \\ \vspace{4pt} 
 \frac{1}{2} & \frac{1}{2} & \frac{1}{2} & 0 & 0 & \frac{1}{2}
\end{pmatrix}~ .
\label{kijparent}
\end{equation}

By and large, the set of $V_{i}$ vectors determines a corresponding symmetry-breaking pattern following
the generic scheme as
outlined in Sect.~\ref{subsectA}.~
In this case, 
$V_0$ and $V_1$ lead to a modular-invariant 
${\cal N}=4$~SUSY theory (in 4D).  $V_2$ then induces a breaking to ${\cal N}=2$~SUSY and an $SO(16)$ gauge group, 
while $V_5$ and the orbifold action $b_3$ induce a further breaking to ${\cal N}=1$~SUSY and $SO(10)$. 
Finally
$V_7$ induces a breaking to $SU(5)$. 
The action $b_{3}$ accompanies the compactification down to 4D on a freely acting orbifold $\mathbb{T}_2/\mathbb{Z}_2$.
This introduces a twist around the $\mathbb{T}_{2}$, through a $U(1)$ subgroup of the internal  symmetry [$SO(4)$ in 
our construction] associated with the compactification from 10D to 6D.~ 
As in Ref.~\cite{Abel:2015oxa}, this structure is loosely based on the 4D MSSM-like models of 
Refs.~\cite{Antoniadis:1990hb, Faraggi:1991be, Faraggi:1991jr, Faraggi:1992fa, Faraggi:1994eu, 
Dienes:1995sv, Dienes:1995bx}.

Given the specifying data in Table~\ref{table:A} and Eq.~(\ref{kijparent}),
the spectrum of the resulting
string model is uniquely determined.
However, string models typically give rise to large numbers of states,
and lists of these states from both the twisted and untwisted sectors
are generally quite long and not always particularly illuminating.
We have therefore off-loaded these lists into Ref.~\cite{URL}.
Certain features of the resulting spectrum are nevertheless important and can easily be summarized.
In particular, the gauge group of the resulting string model is
\beq
     G_{\rm GUT}~=~ SU(5)\otimes SO(6)^2 \otimes [U(1)]^9~.
\eeq
Likewise, the pseudo-anomalous $U(1)$ in this model turns out to be
\begin{equation}
   U(1)_A~=~-2U(1)_{2}+7U(1)_{5}-U(1)_{6}-U(1)_{7}+8U(1)_{8}+2U(1)_{10}-12U(1)_{\rm Tr} ~,
\label{eq:SU5anomu1}
\end{equation}
where $U(1)_{\rm Tr}$ denotes the trace of the $U(5)$ gauge group.
As always, the 
NS-NS sector gives rise to the gravity multiplet as well as the massless scalar states required to build ${\cal N}=2$ gauge multiplets and hypermultiplets. There are five untwisted $\mathbf{10}$
representations of the $SU(5)$. Along with the $\mathbf{5}\oplus\mathbf{1}$ states
they fill out complete chiral $\mathbf{16}$ spinorial representations of the parent $SO(10)$. In addition we find a single conjugate representation. Thus the GUT
model has a total of {\it four net generations of chiral matter fields}\/.
Indeed, the vector-like pair of {\bf 16}'s is precisely the origin of
the Higgs {\bf 5}-plets alluded to in Eqs.~(\ref{Higgsesreps}) and (\ref{Higgsreps2}).

\begin{table}[H]
\centering
\resizebox{\textwidth}{!}{
\begin{tabular}{||c||*{1}c|{l}||}
\hline
\hline
~{Sector}~& {\small $\psi_{\rm s.t.}^{1}\yyr  \psi_{\rm s.t.}^{2}  \psi^{3~}\yyr y^{3~}\yyr\omega^{3~}\yyr\psi^{4~}y^{4~}\yyr\omega^{4~}$} &{\small $\overline{y}^{34}\yyr\overline{\omega}^{34}
\yyr\overline{y}^{56}\yyr\overline{\omega}^{56}\yyr\overline{\psi}^{1}\yy\overline{\psi}^{2}\yy\overline{\psi}^{3}\yy\overline{\psi}^{4}\yy\overline{\psi}^{5}\yy
\overline{\eta}^{1}\yy\overline{\eta}^{2}\yy\overline{\eta}^{3}\yy\overline{\phi}^{1} \yy\overline{\phi}^{2}\yy\overline{\phi}^{3}\yy\overline{\phi}^{4}\yy\overline{\phi}^{5}
\yy\overline{\phi}^{6}\yy\overline{\phi}^{7}\yy\overline{\phi}^{8}$}
\\ \hline
$V_0$&1\xx 1\xx 1\xx 1\xx 1\xx 1\xx 1\xx 1\xx &1\xxl  1\xxl  1\xxl 1\xxl  1\xxl 1\xxl 1\xxl  1\xxl 1\xxl 1\xxl 1\xxl 1\xxl 1\xxl 1\xxl 1\xxl 1\xxl 1\xxl 1\xxl 1\xxl 1 \\
$V_1$&0\xx 0\xx 0\xx 1\xx 1\xx 0\xx 1\xx 1\xx &1\xxl  1\xxl  1\xxl 1\xxl  1\xxl 1\xxl 1\xxl  1\xxl 1\xxl 1\xxl 1\xxl 1\xxl 1\xxl 1\xxl 1\xxl 1\xxl 1\xxl 1\xxl 1\xxl 1 \\
$V_2$&0\xx 0\xx 1\xx 0\xx 1\xx 1\xx 0\xx 1\xx &0\xxl  1\xxl  0\xxl 1\xxl  0\xxl 0\xxl 0\xxl  0\xxl 0\xxl 0\xxl 1\xxl 1\xxl 1\xxl 1\xxl 1\xxl 1\xxl 1\xxl 1\xxl 1\xxl 1 \\
$b_3$&1\xx 0\xx 1\xx 0\xx 0\xx 0\xx 0\xx 1\xx &0\xxl  0\xxl  0\xxl 1\xxl  1\xxl 1\xxl 1\xxl  1\xxl 1\xxl 0\xxl 1\xxl0\xxl 1\xxl 0\xxl 1\xxl 0\xxl 0\xxl 0\xxl 1\xxl 1 \\
$V_5$&0\xx 0\xx 0\xx 0\xx 0\xx 0\xx 1\xx 1\xx &1\xxl  0\xxl  0\xxl  1\xxl  0\xxl  0\xxl 0\xxl  0\xxl 0\xxl 0\xxl 1\xxl1\xxl 0\xxl 1\xxl 1\xxl 1\xxl 1\xxl 0\xxl 1\xxl 1 \\
$V_7$&0\xx 0\xx 0\xx 1\xx 1\xx 0\xx 0\xx 0\xx &0\xxl  1\xxl  0\xxl  1\xxl  $\frac{1}{2}$\xxh  $\frac{1}{2}$\xxh $\frac{1}{2}$\xxh  $\frac{1}{2}$\xxh $\frac{1}{2}$\xxh $\frac{1}{2}$\xxh $\frac{1}{2}$\xxh$\frac{1}{2}$\xxh 1\xxl 1\xxl 1\xxl 1\xxl 1\xxl 0\xxl 1\xxl 0 \vspace{0.8pt} \\
\hline 
${\bf e}$&0\xx 0\xx 1\xx 0\xx 1\xx 1\xx 0\xx 1\xx &1\xxl  0\xxl  0\xxl 0\xxl  0\xxl 0\xxl 0\xxl  1\xxl 1\xxl 1\xxl 0\xxl 1\xxl 0\xxl 1\xxl 1\xxl 0\xxl 0\xxl 0\xxl 1\xxl 0 \\
\hline
\hline
\end{tabular}}
\caption{~Spin structure of the 4D $\mathcal{N}=0$ model. This structure is accompanied by two bosonic degrees of freedom compactified on a $\mathbb{T}_2/\mathbb{Z}_2$ orbifold with twist action corresponding to the vector $b_{3}$.  For notational simplicity, each entry in this table is to be understood as multiplied by $-1/2$.}
\label{table:B}
\end{table}

Given this GUT model, we can now proceed to construct our final, ${\cal N}=0$ SM-like string model.
The boundary-condition vectors that generate this model are shown in Table~\ref{table:B},
and the $k_{ij}$ phases are the same as 
in Eq.~\eqref{kijparent}. 
The introduction of the vector $\mathbf{e}$ breaks the gauge group down to 
\beq
        G ~=~ G_{\rm visible}\otimes G_{\rm semi-hidden}\otimes G_{\rm hidden}~,
\eeq
where 
\beqn 
         G_{\rm visible} &=& SU(3)_c\otimes SU(2)_L\otimes U(1)_Y~\nonumber\\
         G_{\rm semi-hidden} &=& [U(1)]^{11}~\nonumber\\
         G_{\rm hidden} &=& [SO(4)]^{2}~.
\eeqn
The hypercharge of the SM particles is determined by
\beq
\frac{1}{2}U(1)_Y~=~-\frac{1}{3}\left[U(1)_{\overline{\psi}^1}+U(1)_{\overline{\psi}^2}+U(1)_{\overline{\psi}^3}\right]
    +\frac{1}{2}\left[U(1)_{\overline{\psi}^4}+U(1)_{\overline{\psi}^5}\right]~,
\eeq
and the theory is anomaly-free apart from the pseudo-anomalous $U(1)$ combination
\beq
  U(1)_A ~=~ 2U(1)_{0}- 2U(1)_{2}+7 U(1)_{6}-U(1)_{7}-U(1)_{8}+8 U(1)_{9}+2 U(1)_{13}+2 U(1)_{14}-15 U(1)_{\rm Tr'} ~ ,
\eeq
where $U(1)_{\rm Tr'}$ is the linear combination of the trace of $U(3)$ and $U(2)$, 
descending from the $U(5)$ trace in the GUT theory. 
The entire combination descends from Eq.~(\ref{eq:SU5anomu1}) along with
the components of two extra $U(1)$'s in the $[SO(6)]^2$ factors of the GUT theory.

Just as with our previous GUT-like model, the complete particle content of this model is listed in Ref.~\cite{URL}. 
In general, one finds that the twisted 
sectors remain (globally) supersymmetric whereas in the untwisted sectors all states satisfying $q_e={\mathbf {e\cdot Q}} \neq 0$~mod (1) are lifted. 
The graviton and gauge-boson states are generally identified in the NS-NS sector in the usual way, along with the complex radion and antisymmetric tensor.  These states are massless after the CDC is applied since they
unavoidably have $q_e=0$.  Conversely, the gravitino as well as the gauginos become massive after the CDC.~ We should add that it is possible for linear combinations of the basis vectors 
$\{V_0,\dots,V_7\}$ to produce sectors that yield additional gauge bosons in the spinorial representations of the observable $SU(3)\otimes SU(2)$ and/or hidden gauge groups, indicating unwanted potential gauge 
enhancement. One has to be careful to ensure that such states are indeed projected from the massless physical spectrum of the theory by the generalized GSO projections, a requirement which partly determines the 
structure constants. 

As mentioned in the general discussion of Sect.~\ref{entwined},
complete anomaly cancellation requires the addition of extra orbifold twisted sectors that have an ${\bf e}$ action. As shown in Ref.~\cite{Aaronson:2016kjm}, such sectors 
must be present because ${\bf e}$ simply becomes another 
projection vector in the small-radius limit.
These sectors typically provide extra hidden states that ensure consistency; 
moreover, being twisted, these states are supersymmetric and 
cannot gain any radius-dependent breaking from the CDC.~ 
Moreover, it was shown in Ref.~\cite{Aaronson:2016kjm} that the small-radius limit corresponds to structure constants of the form $k_{ei}=0$ and $k_{ee}=1/2$, and that there 
exist further unexplored possibilities that have non-trivial structure constants. For the present discussion, the easiest way to determine the additional twisted 
sectors is to add the vector $V_8={\bf e}$ to the theory, along with the choices 
$k_{8i}=0$ and $k_{88}=1/2$. This trick works because this vector cannot generate new massless untwisted sectors involving $V_8$ since the 
GSO condition for a sector $\alpha V\equiv V_8 + \alpha_i V_i $ is given by $V_8\cdot {\bf Q} = 1/2$, which inevitably conflicts with the CDC-shifted Virasoro operators. In fact 
the untwisted states that remain are precisely those odd winding modes that will {\it become}\/ massless 
in the zero-radius limit, whereas the desired twisted states of the zero-radius limit must already be present
and massless.

It turns out that this model gives rise to eighteen sets of Higgs pairs which include states of the generic form shown in Eq.~(\ref{Higgsesreps}). 
Explicitly, in the notation of Ref.~\cite{FF3}, the Higgs states remaining in the NS-NS sector 
({\it i.e.}\/,  the $\mathbf{0}$ sector) are given by
\begin{eqnarray}
h_{u}^{(1),(2)}&=& \{b,d\}_{\rm s.t., -\frac{1}{2}}^{2}|0\rangle_R\otimes \tilde{b}^{4,5}_{-\frac{1}{2}} \:\tilde{d}^{1}_{-\frac{1}{2}}|0\rangle_L \nonumber \\
h_{d}^{(1),(2)}&=&\{b,d\}_{\rm s.t., -\frac{1}{2}}^{2}|0\rangle_R\otimes \tilde{d}^{4,5}_{-\frac{1}{2}} \:\tilde{b}^{1}_{-\frac{1}{2}}|0\rangle_L \, .
\end{eqnarray}
The other Higgses as well as the singlets are produced from various untwisted or twisted sectors, and they all carry charges under the semi-hidden sector gauge group. It is worth emphasizing that
although the supersymmetric counterparts of the Higgses and the singlets are lifted, there are Higgsino states as well as singlino states which are not lifted by the CDC and
which have different horizontal charges. Such states fill out $e$-SUSY multiplets for the reasons outlined 
in Sect.~\ref{eSUSYintro}.~
However, 
as we will see in Sect.~\ref{pheno},
these states can be lifted by their
Yukawa couplings.

As anticipated in our general discussion,
once the $SU(5)$ GUT symmetry is broken by the CDC there are only two complete chiral fermion generations that survive from the original four net supermultiplets of the GUT.~  One generation is in the 
$\overline{V_0+V_2}$ sector and the other is in the $\overline{V_0+V_1+V_2+\alpha_7 V_7}$ sector. 
Although there are ultimately no superpartners for the chiral matter fields from the untwisted sector, it can be verified 
that the spectrum exhibits $e$-SUSY, as expected,
with smatter fields bearing different horizontal charges. 
Indeed one can identify the charge vector $ {\bf Q}_{\rm horiz}$ in Eq.~(\ref{eq:qhoriz}) that distinguishes a field and its $e$-partner. For example  
the field $d^{c \scriptstyle (4)}$ occurs in the $\overline{V_0+V_2+V_5+ V_7}$ sector. Its $e$-partner   $\tilde{d}^{c \scriptstyle (12)}$ appears in the $\overline{V_1+V_2+V_5+ V_7}$ sector, which is also where its 
true superpartner appears in the theory 
when there is no Scherk-Schwarz twist.  
Inspecting similar $e$-partner pairs, we find that $ {\bf Q}_{\rm horiz}=U(1)_0\oplus U(1)_{11}$.

Overall, this particular model has a total of $N_b^{(0)} =552$ complex bosonic degrees of freedom at the massless level.
Most importantly, however, it also has $N_f^{(0)} =552$ complex fermionic degrees of freedom at the massless level.
Thus the one-loop cosmological constant is exponentially suppressed, making this the first construction of a 
{\it metastable, non-supersymmetric}\/ SM-like theory which also incorporates the all-important 
GUT-precursor and {\it e}\/-SUSY structures discussed in Ref.~\ref{entwined}.~
As such, the existence of models of this type then paves the way for genuine phenomenological model-building.


\section{Additional phenomenological aspects of metastable string models}
\label{pheno}

In this section we briefly comment on two additional phenomenological aspects that are important for general non-SUSY 
metastable string models:
scalar fields acquiring large VEVs, and 
the ever-present issue of proton decay.
As we shall see, both of these issues have interesting resolutions within our metastable string models ---
resolutions which are directly connected to the GUT-precursor and {\it e}\/-SUSY structures we have been discussing.

\subsection{Large VEVs and the existence of an approximate moduli space}

It is important that the scalar fields in our model be able to accrue large VEVs --- 
 {\it i.e.}\/, VEVs which are significantly larger than the weak scale.
One reason for this is that such  VEVs are a feature of Green-Schwarz (GS) anomaly cancellation. 
Issues surrounding the GS mechanism in our scenario will be discussed further below.
A second motivation for considering large scalar VEVs is purely phenomenological:
given that the renormalisable superpotential couplings are all degenerate, 
introducing flavor structure into the Yukawa couplings 
will typically require additional VEVs for so-called flavons.   
Finally, large scalar VEVs are also needed 
in order to lift the masses of many of the matter fields that would otherwise appear in the 
KK spectrum below $M_{\rm GUT}$ 
and thereby produce the negative universal beta-function coefficient $\tilde b<0$ required within the 
GUT-precursor paradigm.

However, these observations then raise a generic question:
is it possible to give large VEVs without changing the
conclusion that the dilaton tadpole is suppressed?
In general,
such VEVs would be expected to lift 
the cosmological constant unless there is a residual flatness in the potential. 
Thus, in order to be able to assign large VEVs to the scalar fields in our model 
without destroying the stability of the model itself, we must require that the model exhibit 
an  {\it approximate moduli space}\/ --- 
{\it i.e.}\/,  a space of almost flat directions that 
remains even after the breaking of supersymmetry.

We begin by discussing Green-Schwarz (GS) anomaly cancellation, as this issue is critical for our analysis.
Generally in heterotic theories, and in the models presented in the previous section, there is a single 
anomalous $U(1)_A$
symmetry whose mixed anomalies are cancelled by the Kalb-Ramond field. 
At the level of the effective theory, the spontaneous
breaking of SUSY can be understood within an entirely ${\cal N} = 1$ supergravity formulation as a linear complex-structure modulus term
in the superpotential, as described in Ref.~\cite{Abel:2016hgy}.
Of course,
since the gravitino mass in our theory is of the same order as the masses of the lowest-lying KK modes
regardless of the size of the compactification radii,
our theory is never really described by a 4D supergravity; 
indeed,
a 4D supergravity treatment is ultimately not useful for describing
quantities such as scattering amplitudes.    However, such a treatment is nevertheless useful 
in the sense that 
it successfully reproduces the vacuum structure
and spectrum of the low-energy theory.
The Green-Schwarz mechanism then corresponds to the generation of  a Fayet-Iliopoulos term in this effective theory, which in turn leads to a VEV for one or more of the scalar fields appearing in the anomalous $D_A$-term~\cite{DSW}.
Typically one expects such VEVs to be  string-sized, while the masses of any fields that couple to those scalars are lifted. 

In this connection it is important to stress that since the cosmological constant is being calculated directly in a non-supersymmetric string theory, 
this effective GS field theory picture is inferred ``after the fact''  of direct computation.  In other words, the 
exponentially suppressed one-loop cosmological constant that 
one calculates in the string theory is the value that arises
when one {\it is already sitting in the correct vacuum}. This is analogous to the fact that the anomalous photon is found to have a non-zero mass within string perturbation theory, and that in a supersymmetric theory one finds a one-loop cosmological constant that is precisely zero.  These are values that arise in the correct (shifted) vacuum of the supergravity theory, but not in the original (unshifted) vacuum.

It is also important to note that the GS mechanism in our scenario is not typical:  the associated scales are adjusted from the usual ones by the large volume.  In general, the potential is minimized where the anomalous $D$-term vanishes, which in turn implies a constraint of the form 
\beq
    \sum_i q_{X,i} |\hat{\phi} |^2 ~\sim ~\xi  g_4^2 M_s^2~,
\label{GSscales}
\eeq
where $q_i$ are the $U(1)_A$ charges, 
where $\xi =  {\rm Tr}\,( q_X)/(192\pi^2)$ is the anomaly coefficient, 
where $\hat{\phi}_i$ are the canonically normalized fields, and where $g_4$ is the tree-level coupling 
(without threshold corrections).  However, as we have seen in Sect.~\ref{2a},
the coupling $g_4$ is suppressed by the compactification volume.
Indeed, when there are $\delta=2$ large orthogonal dimensions of radius $R$, we learn from 
Eq.~(\ref{scalerelation}) that
\beq
        g_4 ~=~ {M_s\over M_P} ~=~ \sqrt{{ -16\pi \over \tilde b}} \,{1\over M_s R}~.
\label{fivetwo}
\eeq
Eq.~(\ref{GSscales}) then becomes
\beq
    \sum_i q_{X,i} |\hat{\phi} |^2 ~\sim ~ - {{\rm Tr}\, (q_X)\over 12\pi \tilde b} M_{\rm KK}^2~,
\eeq
from which we see that the solutions for the VEVs $\langle \hat \phi\rangle$ 
of the canonically normalized scalar fields $\hat \phi$
generally scale as $M_{\rm KK}$ rather than $M_s$.
Indeed, this is true even though the VEVs of the scalar fields  
in the string frame are string-sized.
Thus, if we wish the GS mechanism to retain its usual field-theoretic interpretation, we need only require that
the 4D supergravity description hold up to the KK scale rather than the string scale.  Of course, 
the KK scale is also precisely the scale above which the 4D description breaks down.
Moreover, as we have seen, this scale may be as high as $10^{14}$~GeV.

Given these observations, 
we may now return to our original question, namely that of determining
the conditions under which there exists an approximate moduli space
which would allow such scalar VEVs without disturbing the exponential suppression of the dilaton
tadpole that occurs when $N^{(0)}_b=N^{(0)}_f$. 
Note
that there are two kinds of scalars that we might consider:  those that descend from vector multiplets, and those that originate in 6D hypermultiplets. The former appear in the NS-NS sector of the theory and take the form of off-diagonal (non-abelian) Wilson lines. 
Examples within the model presented in Sect.~\ref{section4}
include the SM singlet fields $\tilde{x}^{(1)}$, $\tilde{x}^{(2)}$, $\tilde{x}^{(11)}$, and $\tilde{x}^{(12)}$ 
as well as  the two generations of Higgs fields, $h_d^{(1,2)}$ and $h_u^{(1,2)}$. 
By contrast, the fields that descend from hypermultiplets 
appear in other ``twisted'' sectors (where ``twisted'' here refers to sectors in the 6D theory, not to the final orbifold compactification to 4D). 
Examples within the model presented in Sect.~\ref{section4}
include $\tilde{x}^{(93)}$ and the remaining Higgses. 
In both cases, the question of flatness is of course synonymous with requiring a radiative 
correction to their squared masses that is much smaller than the generic $1/R$ value, 
which in the present context could be as high as $10^{14}$\,GeV.  

Remarkably, there are already dramatic cancellations in the radiative corrections to the squared masses of fields such as $\tilde{x}^{(1,2)}$ that descend from vector multiplets.  This is ultimately a consequence of the 
$e$-SUSY which is exhibited by the Yukawa couplings, as discussed in Sect.~\ref{entwined}.~ 
In particular, such fields couple equally to a field and its $e$-SUSY partner.  For this reason, $e$-multiplets 
are not able to contribute in the quadratic divergences of their squared masses in the effective field theory. 
In fact, as will be discussed in detail in Ref.~\cite{upcoming}, it turns out that these contributions are 
actually exponentially suppressed in the full string theory.
Note that this cancellation is not a supersymmetric one, since the multiplets involved 
in the cancellation are not supersymmetric partners of each other.
These cancellations nevertheless serve to remove those contributions to the squared masses of these particles
that in the effective field theory would be quadratically divergent, 
much as in the spirit of folded SUSY~\cite{Burdman:2006tz}.

We can see this explicitly within the example model presented in Sect.~\ref{section4}.~
Towards this end,
let us consider raising the masses of 
vector-like pairs of certain superfluous fields (such as the large number of vector-like pairs of down quarks and squarks  and Higgses and leptons appearing in the spectrum tables) through their Yukawa couplings to singlet fields, in what are essentially generalizations of the NMSSM ``$\mu$-term''. 
Given our effective spontaneously-broken classical supergravity description for the light spectrum,
we can think in terms of the relevant pieces of the full superpotential. 
In particular, as dynamical ``$\mu$-terms'' for vector-like matter,
we find~\cite{URL}
\begin{eqnarray}
   \frac{1}{\sqrt{2} \gym }\,    W_{f} & ~\supset ~&  
{\tilde{x}}^{\scriptstyle (1)}\,\left( 
   d^{c \scriptstyle (4)}d^{\scriptstyle (2)}
   +d^{c \scriptstyle (9)}d^{\scriptstyle (8)} +  
\ell^{c \scriptstyle (3)}\ell^{\scriptstyle (3)}+\ell^{c \scriptstyle (5)}\ell^{\scriptstyle (8)} ~+~ 4\times x\mbox{-pairs} \right) 
       \nonumber \\ 
   &&~~+~ 
 {\tilde{x}}^{\scriptstyle (2)}\,\left( 
   d^{c \scriptstyle (3)}d^{\scriptstyle (3)}+d^{c \scriptstyle (10)}d^{\scriptstyle (7)}+ \ell^{c \scriptstyle (2)}\ell^{\scriptstyle (4)}+\ell^{c \scriptstyle (6)}\ell^{\scriptstyle (7)}~+~ 4\times {x}\mbox{-pairs}\right) ~.
 \end{eqnarray}  
By contrast, 
the dynamical ``$\mu$-terms'' for the $e$-SUSY partners of this vector-like matter
are given by
\begin{eqnarray}
   \frac{1}{\sqrt{2} \gym}\,    W_{b} &~\supset ~ &  
   {\tilde{x}}^{\scriptstyle (1)}\left(  
   \tilde{d}^{c \scriptstyle (12)}\tilde{d}^{\scriptstyle (11)}+
 \tilde{d}^{c \scriptstyle (7)}\tilde{d}^{\scriptstyle (6)}+
   h_u^{ (11)}h_d^{ (9)}
+    h_u^{ (16)}h_d^{ (18)} ~+~ 4\times \tilde{x}\mbox{-pairs} \right)
    \nonumber \\
    && ~~+~   
{\tilde{x}}^{\scriptstyle (2)} \left(
   \tilde{d}^{c \scriptstyle (14)}\tilde{d}^{\scriptstyle (13)}
   +\tilde{d}^{c \scriptstyle (5)}\tilde{d}^{\scriptstyle (4)}
   +h_u^{ (9)}h_d^{ (11)}
    +h_u^{ (18)}h_d^{(16)) } ~+~ 4\times \tilde{x}\mbox{-pairs}\right)
 ~ .
   \end{eqnarray} 
The meaning of $W_b$ and $W_f$ is as before, but here we explicitly label the pairs of scalars or fermions that remain light.  Note also that we label all scalar doublets as Higgses.
As expected, $\tilde{x}^{(1)}$ and $\tilde{x}^{(2)}$ couple degenerately to both quark and lepton pairs {\it and} their squark and slepton/Higgs $e$-partners.
On the other hand, this degeneracy is broken for fields such as 
${\tilde{x}}^{\scriptstyle (93)}$ that do not descend from 
gauge multiplets.  Indeed, upon inspection, we find that
${\tilde{x}}^{\scriptstyle (93)}$ couples only to the $d^{c \scriptstyle (2)}d^{\scriptstyle (4)}$  fermions and the scalar doublets from a single $e$-partner parent $SU(5)$ supermultiplet.

We see, then, that certain directions such as ${\tilde{x}}^{\scriptstyle (1,2)}$ ---  which must couple in the GUT theory to both superfield parents of an entwined pair ---  will, upon acquiring a VEV,  
give degenerate masses to the entire $e$-multiplet.  
Thus the boson/fermion degeneracy in the massless sector can only be affected by their couplings to gauge fields. 
Unfortunately, since even SM singlets such as ${\tilde{x}}^{\scriptstyle (1,2)}$ are charged under horizontal $U(1)$'s, they give masses to gauge bosons and the cancellation of the quadratic divergences is not generally complete. 
In particular, even though the $U(1)$ symmetries that are not broken by the Scherk-Schwarz compactification are 
phenomenologically
global symmetries, without {\it complete}\/ cancellation singlets such as 
${\tilde{x}}^{\scriptstyle (1,2)}$ 
will acquire masses that essentially stem from Eq.~(\ref{fivetwo}).
Thus, the squared masses induced by couplings to gauge fields will have general magnitudes
\beq
    m^2_{{\tilde{x}}^{\scriptstyle (1,2)}} ~\sim~ {g_4^2\over 16\pi^2}\,M_s^2 ~\sim~ {1\over 16\pi^2 R^2}~,
\eeq
and large VEVs would induce a correspondingly large cosmological constant. 
Indeed, complete cancellation and exponentially flat singlet 
directions would require coupling to an equal number of fermion degrees of freedom that correspond to off-diagonal gauginos that are left light by the Scherk-Schwarz/GUT breaking. 
Furthermore, since the anomalous linear combination of the gauge bosons is massive due to the GS mechanism, such 
considerations
involve more than simply counting massless tree-level degrees of freedom.  

Ultimately, the best way to address this issue is through a calculation in the full string theory.
This will be discussed in detail in Ref.~\cite{upcoming}. 
The upshot of that calculation is that ensuring the cancellation of the leading squared-mass contributions  simply becomes a matter of correctly adjusting the breaking of gauge symmetry by the Scherk-Schwarz phases. In particular, there are no tuneable couplings: either a theory ends up with the correct particle content to produce light scalars or it does not.  Moreover, it turns out that there is a high degree of degeneracy in the squared masses that are radiatively induced in this manner: if the leading contribution cancels for one scalar descending from a higher-dimensional gauge multiplet, it is likely to cancel for them all. Thus we indeed expect certain theories to enjoy large approximate moduli spaces, as desired.

If such a configuration can be found, then
the  structure of the spectrum of such string models 
remains as originally described in Sect.~\ref{sec:framework}
regardless of the VEVs of fields like  $\tilde{x}^{\scriptstyle (1,2)}$. 
Indeed, each KK level all the way up to the first string excitations 
remains boson/fermion degenerate. 
In other words, such directions in field space remain almost flat
  in the presence of one-loop corrections. 
By contrast, heavy directions such as ${\tilde{x}}^{\scriptstyle (93)}$ would lift different numbers of bosons and fermions, and therefore the cosmological constant will receive large corrections if such scalars acquire a VEV. 
In this connection, we note that
that VEVs whose magnitudes are integer multiples of $1/R$ can result in a restoration of boson/fermion degeneracy, 
KK level by KK level, even for heavy fields.  Indeed, such directions are actually {\it periodic}\/ 
in nature due to spacetime modular symmetries.
This is a familiar phenomenon that can also be seen directly if these directions are identified with a continuous Wilson line, as in Ref.~\cite{Abel:2016hgy}.

Finally, of course, we remark that we ultimately wish at least one of the Higgs fields to correspond to such a flat direction as well.
Likewise, we seek to have most of the scalars acquire squared masses of order  
$\sim 1/16\pi^2 R^2$, given by the scale of SUSY breaking in the usual manner. 
This will be  discussed elsewhere~\cite{upcoming}.

\subsection{Yukawa couplings and the general outlook for proton decay}

Another primary issue of interest in such theories 
is the identification of matter, in particular the generation assignment, and the resulting Yukawa couplings. 
This then naturally leads to the question of proton decay. 

Towards this end, we begin by presenting the Yukawa couplings for our example model. 
The renormalizable (dimension-three) non-vanishing Yukawa couplings, written as superpotential terms, can be determined in a relatively straightforward manner, and the results for our model
are listed in Ref.~\cite{URL}. 
They include terms such as $W_{\rm Yuk} \supset 
h_d^{ (3)}q^{\scriptstyle (2)}d^{c \scriptstyle (5)}$ but for the specific  model presented here do not include top Yukawa couplings at leading order. No dimension-four Yukawa couplings for the matter fields are allowed by the horizontal charges. The leading-order Yukawa couplings for the up-quarks come from dimension-five operators involving SM singlet fields $\tilde{x}^i$.  These couplings would obviously require a boson/fermion degeneracy-preserving VEV to be generated for the singlets in the manner discussed above.
The down-quark mass matrix and electron mass matrix are both rank two, 
and thus the end result can be two light generations with filled Yukawa couplings.

Finally we comment on proton decay. In general, it can be subtle to determine whether the first generation of chiral matter fields should be twisted or untwisted.  
However the absence of rapid proton decay suggests that in $e$-SUSY scenarios,
the first generation {\it cannot} be twisted.
To see this, let us consider for example the dimension-six proton-decay operators:
\begin{equation}
\frac{qqql}{\Lambda^2}~, ~~~~ \frac{d^c u^c u^c e^c}{\Lambda^2}~, ~~~~ 
 \frac{\overline{e^c}\overline{u^c}qq}{\Lambda^2}~, ~~~~ \frac{\overline{d^c}\overline{u^c}ql}{\Lambda^2}~,
\end{equation}
where $\Lambda \sim {\cal O}(M_{\rm GUT})$ and where clearly the states are all first generation. 
The discussion in Sect.~\ref{eSUSYintro} tells us that the twisted sectors of these theories must fall 
into complete SUSY GUT
multiplets;  indeed the twisted sectors are insensitive to the CDC and the spectrum is entirely unaffected by it. 
Therefore all of the above operators are  uncharged and one would expect them to be generated in the 
Lagrangian at the same order as in usual SUSY GUTs but with $\Lambda \sim 1/R$. 
As a consequence of the hierarchy $1/R \ll M_{\rm GUT}$, such operators would 
automatically lead to disastrously rapid proton decay. By contrast, the states in
 {\it untwisted} generations are {\it not}\/ in GUT multiplets. 
Consequently, the above proton-decay operators all carry a non-zero horizontal charge.
Such operators therefore cannot be generated in the Lagrangian, and 
the proton therefore cannot decay because this process would violate the conservation of the horizontal charge.


\section{Discussion\label{discussion}}

In Ref.~\cite{Abel:2015oxa} we established a method of constructing non-super\-sym\-metric string models 
that are essentially stable, with near-vanishing one-loop dilaton tadpoles and cosmological constants.  
This then opens up the tantalizing possibility of realizing stable string models whose low-energy limits directly resemble the Standard Model rather than one of its supersymmetric extensions.  In this paper 
we investigated the phenomenological structure of such strings.
Because our construction necessarily involves large-volume compactifications,
one pressing issue concerns the behavior of the gauge couplings.
As we discussed, this then requires that our strings exhibit a variant of the so-called ``GUT precursor'' structure
originally proposed in Refs.~\cite{Dienes:2002bg,Dienes:2004rt}.~
Tightly coupled with this, we also found that the spectra of such strings exhibit a so-called ``entwined SUSY''
(or {\it e}\/-SUSY) in which members of the same $e$-multiplet have different charges under a 
horizontal $U(1)$ symmetry.  Thus, this horizontal $U(1)$ symmetry is non-trivially ``entwined'' with
the same physics that renders the theory non-supersymmetric and also breaks the GUT symmetry.

We then proceeded to construct an actual SM-like heterotic
string model which provides direct illustration of these observations.
This model is non-supersymmetric and yet displays 
both the GUT-precursor and {\it e}\/-SUSY structures,
all while remaining one-loop stable with a near-vanishing one-loop cosmological constant.
We then discussed some general features of such models, in particular how it is possible to give 
masses to unwanted states and how singlet VEVs can be accommodated, all without ruining the stability properties. 
Specifically, we found that the contributions 
from all {\it e}\/-multiplets 
actually cancel.    
Thus {\it e}\/-SUSY implies a remarkable cancellation of all purely matter loop
contributions to scalar squared masses at one loop, which in turn means
that instability at large VEVs is a function only of the pattern of the {\it gauge}\/ breaking.
Consequently, if there are models for which these contributions vanish as well, the corresponding cosmological constant 
would be exponentially
insensitive to such VEVs.
Indeed, this might be the basis upon which one might hope 
to build a complete phenomenology.
We also found that the horizontal $U(1)$ symmetries that are required in such models
naturally also protect against the kinds of rapid proton decay that might otherwise occur 
in such GUT-precursor models.

One striking observation concerning our results --- undoubtedly connected with the cancellations 
discussed above --- is that $e$-SUSY, which emerges automatically
in our construction,  has a structure which bears a strong resemblance 
to the structures underlying ``folded SUSY''~\cite{Burdman:2006tz},  ``supersymmetry in slow motion''~\cite{Chacko:2008cb}, and more recently ``hypertwisted SUSY''~\cite{Kats:2017ojr}.  Indeed all of these latter symmetries 
have been proposed in the literature as mechanisms for stabilizing the weak scale 
in a way that might be hard or impossible to detect at current colliders.
Of course such ``neutral naturalness'' was never the purpose of {\it e}\/-SUSY.
It is nevertheless possible that relatively slight modifications of the construction we have developed here
might be capable of yielding either folded SUSY or one of its variants --- all while remaining within our
overall string framework which ensures a near-vanishing one-loop dilaton tadpole.
Indeed, such a construction would undoubtedly continue to involve GUT precursors and the emergence 
of a UV fixed point, as discussed in Sect.~\ref{precurs}.~
Such a construction 
would thus provide a useful and genuinely UV-complete laboratory in which such neutral-naturalness ideas 
might be studied. 

Even within the class of {\it e}\/-SUSY models we have discussed here, however, 
it is important to realize that there might exist a subset having exponentially-suppressed one-loop squared Higgs masses and cosmological constants, simply as a result of a particular choice of particle content.
Indeed, such models may even display accidental discrete symmetries, perhaps realizing one of the neutral-naturalness 
scenarios mentioned above or one which is entirely new.
Moreover, the fact that we are working
within a full string-theoretic framework rather than within a mere effective field theory
already guarantees one important feature:
as discussed in Sect.~\ref{sec:framework},
the fundamental symmetries that underpin the finiteness of generic closed strings
yields severe UV constraints. 
Specifically, modular invariance and 
 misaligned supersymmetry~\cite{missusy,supertraces} 
ultimately restrict the spectrum of the 
theory at all energy scales simultaneously and thereby ensure a desirable UV behavior.
There are consequently almost no arbitrary parameters, and such cancellations would be natural in the sense
that they depend only on the very restricted, discrete choices of particle content.
This issue will be explored in more detail in Ref.~\cite{upcoming}.

Throughout this paper, we have described the string models in our construction
as stable or metastable.    As we have emphasized throughout, this refers to the 
critical issue of dilaton stability --- {\it i.e.}\/, the suppression of the dilaton tadpole (cosmological constant).
The issue of the dilaton tadpole is uniquely problematic in the construction of non-supersymmetric strings, 
as the existence of such a tadpole is the direct hallmark of the breaking of supersymmetry.
Moreover, the dilaton is the only field that cannot easily be given a mass by turning on a VEV.~ 
However, just as for supersymmetric strings, there nevertheless remain further moduli which also require 
stabilization through either string-theoretic or field-theoretic means.
Our models are therefore not fully stable in this sense. 
However, by successfully giving rise to vanishing dilaton tadpoles, 
these models are now on 
essentially equal footing with their more traditional supersymmetric
counterparts.  
These models can then provide the context for the development of 
a non-supersymmetric string phenomenology which is on a par with that of strings 
with spacetime supersymmetry.

In a similar vein,
given that the analysis in this paper has focused primarily on the suppression of the one-loop dilaton tadpole,
another immediate question concerns the behavior that might emerge at higher loops and its possible ramifications
for the ultimate stability of these strings.
In particular, even though the one-loop dilaton tadpole is exponentially suppressed in our construction,
there are likely to exist artificial uncancelled 
divergences at two loops.  This indicates that although our assumption of a Minkowski background is 
a reasonable approximation, the theory is not quite stabilized in this sense either
and thus is not quite in its true vacuum. 
Of course,
due to the exponential suppression of the one-loop tadpole,
 one might assume that the true dilaton-stabilized vacuum is not very far away in field space.  Moreover,
once the theory is stabilized in the true vacuum, 
such higher-loop divergences should cancel as well.

Proceeding further,
one would obviously like to engineer this final stage of stabilization within the context of the full 
string theory so that one can continue to benefit from the UV-complete nature of the theory. 
There are then two positions one can adopt as to what might happen to the cosmological constant. 
First, it is possible that in the hypothetical true vacuum, the two-loop tadpole also cancels 
(or is significantly suppressed)
at leading order. 
In this case
the leading one-loop cancellation would be telling us something profound about the stability of the 
whole theory.  Indeed, it is conceivable that the GUT-precursor and {\it e}\/-SUSY structures 
that remain in the massless sector after one-loop 
stabilization play an important role in helping to ensure
such a result at higher-loop order. 

The second possibility is that
the cosmological constant in the     hypothetical true vacuum
has a generic magnitude consisting of a two-loop factor times $R^{-4}$. In this case 
one could potentially derive a set of additional conditions beyond bose/fermi degeneracy that must be satisfied
 if such two-loop contributions are also to vanish or experience a significant suppression.
Such conditions were derived in Ref.~\cite{Abel:2017rch}, and interestingly 
the condition for the suppression of a generic two-loop cosmological constant is that 
the sum over the one-loop scalar squared masses also be exponentially suppressed. 
This would then suggest a possible connection between the exponential suppression of scalar squared masses and 
higher-loop stability.  Note that this second approach essentially sets us along a 
road originally outlined in Ref.~\cite{Jack:1989tv}, where stability imposes ever more constraints at higher loops until a sufficiently high order is reached. One should bear in mind, however, that each constraint is simply a function of the particle content, and in string theory  we have no possibility for tuning this:  either there exist non-supersymmetric theories that simultaneously satisfy these constraints, or there do not.

At this stage, many of these ideas are somewhat speculative and further work in these directions is needed.  However, the framework we have been investigating here --- and the generic predictions of phenomenological GUT-precursor and {\it e}\/-SUSY structures in the resulting spectrum --- are likely to prove critical in any such analysis.  We can therefore hope that these structures may help us not only in understanding the phenomenology of these strings but also in addressing  the central question of whether the world in which we live can indeed ultimately be viewed as the direct low-energy limit of a stable, non-supersymmetric string.  In fact, despite the string-theoretic nature of this paper, many aspects of our general construction 
can even be formulated and understood within the framework of ordinary higher-dimensional quantum field theory without reference to string theory as the UV completion~\cite{upcoming}.  As such, the overall phenomenology we have outlined here can be viewed as very general expectation which can serve as paradigm for {\it any}\/ UV-complete theory that is rendered non-supersymmetric by Scherk-Schwarz compactification.  Indeed, any such theory would have a leading radius-dependent instability that can be cured by arranging the bosonic and ferminionic KK states as we have described.  The underlying supersymmetry then ensures (by the geometric arguments in the text) that further contributions from the UV completion must be exponentially suppressed at any reasonably large radius.  Moreover, following the logical line in Sect.~\ref{entwined}, the GUT-precursor structure is then a prediction of our need to have phenomenologically viable gauge couplings, as originally described in field-theoretic terms in Refs.~\cite{Dienes:2002bg,Dienes:2004rt}, and in turn the {\it e}\/-SUSY structure follows directly from the consequent entwining of the GUT- and supersymmetry-breaking.  Indeed, it would be interesting to formulate a paradigmatic {\it e}\/-SUSY extension of the Standard Model within a  bottom-up extra-dimensional field theory framework~\cite{upcoming}, following an approach 
similar to that in Ref.~\cite{Garcia:2015sfa}.
We therefore expect that the structures we have outlined here may serve as generic predictions not only of stable, non-supersymmetric strings but indeed of large classes of   non-supersymmetric UV-complete field theories as well.


\vspace{1cm}

\noindent {\bf Acknowledgements}: SAA would like to thank 
the Ecole Polytechnique and the CERN Theory Division for hospitality and support during the completion of this work. 
KRD also thanks the Ecole Polytechnique for hospitality.
SAA acknowledges the Royal Society for support under CostShare grant RS060243. 
The research activities of KRD were supported in part by the U.S.~Department
of Energy under Grant DE-FG02-13ER41976 (DE-SC0009913)
and also in part by the U.S.~National Science Foundation through its employee IR/D program.
SAA and KRD would like to thank Emilian Dudas, Alberto Mariotti, and Herv\'e Partouche for many interesting 
discussions. 
The opinions and conclusions expressed herein are those of the authors, and do not represent any funding agencies.

\vspace{1cm}


\appendix

\section{~More general forms of CDC}
\label{appendix: 1}

In this Appendix, we further consider the technical question 
prompted by the analysis in 
Sect.~\ref{section4}, namely 
the question of how
a complex $1/4$ phase ({\it i.e.}\/, $\psi\rightarrow e^{i\pi/2}\psi=i\psi$) can overlap with the CDC vector and the orbifolding $b_{3}$.  In particular, we shall demonstrate that models with more general complex phases can
still be understood (with certain caveats) in a diagonal basis. 
The discussion here builds very much upon that in Ref.~\cite{Abel:2015oxa}. 

We begin by considering the orbifold and basis-vector actions on a single
complex fermion.  We shall refer to these by the basis vector containing them, 
namely $V_{7}$, $b_{3}$, and $\mathbf{e\cdot Q}$ respectively. The required relations of the three actions are 
\begin{equation}
     [b_{3},V_{7}]~=~[V_{7},\mathbf{e\cdot Q}]~=~\{b_{3},\mathbf{e\cdot Q}\} ~=~0~,
\label{eq:rels}
\end{equation}
where we recall from Ref.~\cite{Abel:2015oxa} that the orbifold commutes with the CDC charge 
in order to have consistent mass eigenstates for the KK tower. 

Let us focus first on the $b_{3}$ and $\mathbf{e\cdot Q}$ relation and try to recreate it using only a single complex fermion. This will eventually inform us how to handle $V_{7}$. Recapping briefly 
from Ref.~\cite{Abel:2015oxa}, we may write this relation in terms of its real components as follows. The creation/annihilation operators become
\begin{eqnarray}
\sqrt{2}b & = & \chi_{1}+i\chi_{2}\nonumber \\
\sqrt{2}d & = & \chi_{1}-i\chi_{2}~,
\end{eqnarray}
where for readability we have dropped mode-number subscripts. 
In terms of these components, $\mathbf{e\cdot Q}=\frac{1}{2}\left(b^{\dagger}b-d^{\dagger}d\right)=i(\chi_{1}^{\dagger}\chi_{2}-\chi_{2}^{\dagger}\chi_{1})$, 
while charge conjugation corresponds to $\chi_{2}\rightarrow-\chi_{2}$. To write these actions in the real formalism, we identify two real fermions as $\chi_{\pm}=\frac{1}{\sqrt{2}}(\chi_{1}\pm\chi_{2})$. According
to Ref.~\cite{Abel:2015oxa}, consistency requires pairs of real fermions 
({\it i.e.}\/, $\chi_{\pm}$) to have equal and opposite shifts under $\mathbf{e}$;  we therefore have 
\begin{eqnarray}
&& \mathbf{e\cdot Q}  ~=~ 
      \frac{1}{2}(\chi_{+}^{\dagger}\chi_{+}-\chi_{-}^{\dagger}\chi_{-})  
   ~=~ \frac{1}{2}(\chi_{1}^{\dagger}\chi_{2}+\chi_{2}^{\dagger}\chi_{1}) \nonumber\\
&&     b_3 \begin{pmatrix}  \chi_+\\ \chi_-  \end{pmatrix}
    ~=~ 
 \begin{pmatrix}
    0 & 1 \\ 1 & 0 
  \end{pmatrix} 
   \begin{pmatrix}
     \chi_+ \\ \chi_- 
    \end{pmatrix}~,
\end{eqnarray}
and we can clearly see that these results inherit the correct conjugation properties. The projection from $b_{3}$ will therefore pick out even or odd eigenstates to be physical, and these are precisely functions of 
$\chi_{+}^{\dagger}\pm\chi_{-}^{\dagger}=\sqrt{2}\chi_{1,2}^{\dagger}$. Meanwhile, in the untwisted sector, $\chi_{+}$ states are shifted by $+\frac{1}{2}$ and $\chi_{-}$ states by $-\frac{1}{2}$. This is
required for a consistent projection (because the orbifolding negates the KK and winding numbers) and simultaneously $\mathbf{e\cdot Q}$.~ It is then a convenience in fermionic strings to note that the entire 
partition function is invariant under a reversal of the sign of the $d^{\dagger}d$ term in $\mathbf{Q}$, which becomes $\mathbf{Q_{r}}=\frac{1}{2}(\chi_{1}^{\dagger}\chi_{1}+\chi_{2}^{\dagger}\chi_{2})$.  Therefore
in the real formalism one must instead use shifts in ${\bf e\cdot Q_{r}}=
   \frac{1}{2} (\chi_{+}^{\dagger}\chi_{+}+\chi_{-}^{\dagger}\chi_{-})
  =\frac{1}{2} (\chi_{1}^{\dagger}\chi_{1}+\chi_{2}^{\dagger}\chi_{2})$, 
with both $b_{3}$ and $\mathbf{e}$ written in the same real basis 
as $b_{3}=-\frac{1}{2}\left[\ldots(01)_{r}\ldots\right]$ and $\mathbf{e}=\frac{1}{2}\left[\ldots(11)_{r}\ldots\right]$.

Clearly this ``real-formalism'' trick will not work for fermions with more general complex phases $V_{7}$. 
However it is still convenient to start in the same way by expressing the vectors as actions on pairs
of real fermions. 
In terms of a single pair of fermions $\chi_{1,2}$, the actions of $b_{3}$ and $V_{7}$ would correspond to 
\beq
b_{3} ~\equiv~  \left(\begin{array}{cc} 1 & 0\\ 0 & -1 \end{array}\right)~,~~~~~
V_{7} ~ \equiv ~ \left(\begin{array}{cc} 0 & -1\\ 1 & 0 \end{array}\right)~.
\eeq
There are therefore no commuting actions for $b_{3}$ and $V_{7}$ involving only a single complex fermion.
On the other hand,  it is possible to find a commuting action involving {\it two}\/ complex fermions. 
Let the fermions be $\psi_{12}=\chi_{1}+i\chi_{2}$ and $\psi_{34}=\chi_{3}+i\chi_{4}$ and their conjugates, with $b_{3}$ giving conjugation in both:
\begin{equation}
b_{3}\left(\begin{array}{c}
\chi_{1}\\
\chi_{2}\\
\chi_{3}\\
\chi_{4}
\end{array}\right)~\equiv~\left(\begin{array}{cccc}
1 & 0 & 0 & 0\\
0 & -1 & 0 & 0\\
0 & 0 & 1 & 0\\
0 & 0 & 0 & -1
\end{array}\right)\left(\begin{array}{c}
\chi_{1}\\
\chi_{2}\\
\chi_{3}\\
\chi_{4}
\end{array}\right).
\end{equation}
In order to commute with $b_{3}$, the $V_{7}$ has to operate on the $(1,3)$ pair, or the $(2,4)$ pair, and thus
takes the form 
\begin{equation}
V_{7}~\equiv~\left(\begin{array}{cccc}
0 & 0 & -1 & 0\\
0 & 0 & 0 & -1\\
1 & 0 & 0 & 0\\
0 & 1 & 0 & 0
\end{array}\right) .
\end{equation}
Eigenstates of $V{}_{7}$ with eigenvalue $\pm i$ are therefore linear combinations of $\psi_{13}=\chi_{1}\pm i\chi_{3}$ and $\psi_{24}=\chi_{2}\pm i\chi_{4}$. Note that $b_{3}\psi_{13}=+\psi_{13}$ and 
$b_{3}\psi_{24}=-\psi_{24}$, so the actions of $b_{3}$ and $V_{7}$ are simultaneously diagonal in the $(\psi_{13},\psi_{24}$) basis, with boundary conditions $V_{7}\equiv-\left(\frac{1}{4},\frac{1}{4}\right)$ and 
$b_{3}\equiv-\left(0,\frac{1}{2}\right)$, respectively.

A suitable form of $\mathbf{e\cdot Q}$ that anticommutes with $b_{3}$ and commutes with $V_{7}$ takes the most general form 
\begin{equation}
\mathbf{e\cdot Q}~=~\left(\chi_{1}^{\dagger}\,\chi_{2}^{\dagger}\,\chi_{3}^{\dagger}\,\chi_{4}^{\dagger}\right)\left(\begin{array}{cccc}
0 & a & 0 & b\\
-a & 0 & b & 0\\
0 & -b & 0 & a\\
-b & 0 & -a & 0
\end{array}\right)\left(\begin{array}{c}
\chi_{1}\\
\chi_{2}\\
\chi_{3}\\
\chi_{4}
\end{array}\right)
\end{equation}
for arbitrary coefficients $(a,b)$.  Thus
\begin{equation}
\mathbf{e\cdot Q}~=~a\left(Q_{12}+Q_{34}\right)+b\left(Q_{23}+Q_{14}\right)~.
\end{equation}

The above discussion pertains to the $(\psi_{13},\psi_{24})$ basis in which $b_{3}$ and $V_{7}$ are diagonal but $\mathbf{e\cdot Q}$ is not.  There is, however, an alternative basis in which $V_{7}$ and 
$\mathbf{e\cdot Q}$ are diagonal but $b_{3}$ is not. 
To find this basis, we note two general properties of antisymmetric matrices: 
their eigenvalues are pure imaginary and appear in pairs along the positive and negative imaginary axis.
The former property makes phases of $1/4$ overlapping with values of $\pm\frac{1}{2}$ in $\mathbf{e}$ particularly easy to treat. Indeed for the above we have two degenerate eigenvalues
$\pm i\sqrt{a^{2}+b^{2}}$. It is natural to then choose $b=0,\, a=1/2$, which gives degenerate eigenvalues for $\mathbf{e\cdot Q}$ of $\pm\frac{1}{2}$. The two eigenvectors with eigenvalues $+1/2$ are 
$b_{12}=\chi_{1}+i\chi_{2}$ and $b_{34}=\chi_{3}+i\chi_{4}$, while not surprisingly those with eigenvalue $-1/2$ are the charge conjugates $d_{12}=\chi_{1}-i\chi_{2}$ and $d_{34}=\chi_{3}-i\chi_{4}$. Given the 
above eigenvectors for $V_{7}$, we are now able to identity the orthogonal linear combinations of $\mathbf{e\cdot Q}$ eigenstates that are also eigenstates of $V_{7}$ with eigenvalue $i$. 
These are given by
\begin{eqnarray}
b_{+}~=~b_{13}+ib_{24} & = & \chi_{1}-\chi_{4}+i(\chi_{3}+\chi_{2})\nonumber \\
b_{-}~=~b_{13}-ib_{24} & = & \chi_{1}+\chi_{4}+i(\chi_{3}-\chi_{2})~,
\end{eqnarray}
with the $\pm$ subscript indicating the $\mathbf{e\cdot Q}$ charge.  Likewise, the eigenstates
with  eigenvalue $-i$ are given by
\begin{eqnarray}
d_{+}~=~d_{13}-id_{24} & = & \chi_{1}-\chi_{4}-i(\chi_{3}+\chi_{2})\nonumber \\
d_{-}~=~d_{13}+id_{24} & = & \chi_{1}+\chi_{4}-i(\chi_{3}-\chi_{2})~.
\end{eqnarray}
It is easy to verify, {\it e.g.}\/, that indeed $V_{7}b_{+}=ib_{+}$.  We can also verify that  in this $(\psi_{+},\psi_{-})$ basis, the charges are now diagonal, {\it i.e.}\/, that $\mathbf{e\cdot Q}=\frac{1}{2}\left(b_{+}^{\dagger}b_{+}-d_{+}^{\dagger}d_{+}
-b_{-}^{\dagger}b_{-}+d_{-}^{\dagger}d_{-}\right)=\frac{1}{2}\left(Q_{12}+Q_{34}\right)$. 
Meanwhile $b_{3}$, which sends $\chi_{2,4}\rightarrow-\chi_{2,4}$, is readily identified as the charge-conjugation permutation
$\psi_{\pm}\leftrightarrow\psi_{\mp}$, or in other words 
\begin{equation}
b_{3}~\equiv~\left(\begin{array}{cc}
0 & 1\\
1 & 0
\end{array}\right)~ .
\end{equation}
Thus in the $(\psi_{+},\psi_{-})$ basis we may write the boundary condition and CDC 
phases as $V_{7}\equiv(\frac{1}{4},\frac{1}{4})$, and $\mathbf{Q_{e}}=(\frac{1}{2},-\frac{1}{2})$. 

Let us now consider the physical states that obey all of the above constraints. 
In particular, we shall focus on the massless sector and determine 
how the projections act on states at this particular mass level. States in the post-CDC 
$\mathcal{N}=2$ theory ({\it i.e.}\/, the theory prior to the $b_{3}$ projection) fall into KK/winding towers, 
$|\psi_{q_{e},m,n}^{(\mathcal{N}=2)}\rangle$, where $q_{e}=\pm\frac{1}{2}$ is the $\mathbf{e\cdot Q}$ charge
of the state. The orbifold then projects this spectrum to states that are invariant under reversal of $q_{e},m,n$, along with other possible phase shifts due to the action of $b_{3}$. However, the $b_{3}$ 
projection was carefully chosen to commute with the other projections that resulted in the $\mathcal{N}=2$ theory. 
Thus the invariant eigenstate of the orbifold may be written as 
\begin{equation}
|\Psi_{phys}\rangle~=~\frac{1}{\sqrt{2}}\left[|\psi_{q_{e},m,n}^{(\mathcal{N}=2)}\rangle+(-1)^{\hat{b}_{3}.\mathbf{Q}}|\psi_{-q_{e},-m,-n}^{(\mathcal{N}=2)}\rangle\right]~,
\end{equation}
where $\hat{b}_{3}\cdot \mathbf{Q}$ accounts for the other possible phase shifts induced by $b_{3}$ for this particular state. For example, the state $b_{+}^{\dagger}|\frac{1}{2},m,0\rangle+b_{-}^{\dagger}|-\frac{1}{2},
-m,0\rangle$ has a $V_{7}$ charge of $i$, sits at mass level $(m+\frac{1}{2})/2R$, and is clearly an eigenstate of $b_{3}$, which simply exchanges the two components. 
Meanwhile $b_{+}^{\dagger}|\frac{1}{2},m,0\rangle-
b_{-}^{\dagger}|-\frac{1}{2},-m,0\rangle$ is projected out by $b_{3}$. This implies that the expressions in the untwisted sector of the partition function are independent of the orbifolding. Indeed, as long as 
Eq.~\eqref{eq:rels} holds, the partition function can be written as
\begin{equation}
\mathcal{Z}(\mathbf{e})~=~\frac{1}{2}\left(\mathcal{Z}\left[\begin{array}{c}
0\\
0
\end{array}\right](\mathbf{e})-\mathcal{Z}\left[\begin{array}{c}
0\\
0
\end{array}\right](\mathbf{0})\right)+\frac{1}{2}\left(\mathcal{Z}\left[\begin{array}{c}
0\\
0
\end{array}\right](\mathbf{0})+\mathcal{Z}\left[\begin{array}{c}
g\\
0
\end{array}\right]+\mathcal{Z}\left[\begin{array}{c}
0\\
g
\end{array}\right]+\mathcal{Z}\left[\begin{array}{c}
g\\
g
\end{array}\right]\right).
\end{equation}
Note that the final three terms are orbifold-twisted and thus lack any dependence on ${\bf e}$.
The second bracket can be identified as the partition function of the original $\mathcal{N}=1$ theory. It is independent of the CDC vector $\mathbf{e}$ and thus vanishes. It could be formally evaluated in the basis where
$b_{3}$ is diagonal in the usual way.  By contrast, the first bracket corresponds to the untwisted $\mathcal{N}=2$ contribution and {\it does}\/ depend on the vector $\mathbf{e}$, but its dependence on $b_{3}$ is trivial:  as described above, the factor of $\frac{1}{2}$ is sufficient to encompass the effect of the orbifold projection on the untwisted states.

Thus, due to Eq.~\eqref{eq:rels}, the orbifold in such theories has a rather trivial interaction with $\mathbf{e}$ in the sense that if a state $|\psi_{m,n}\rangle$ exists in the untwisted sector of the original 
theory, then either $q_{e}=0$ (in which case it undergoes the same orbifold projection as in the non-CDC $\mathcal{N}=1$ theory which is incorporated in the second bracket above), or both 
$|\psi_{q_{e},m,n}^{(\mathcal{N}=2)}\rangle$ and its partner state $|\psi_{-q_{e},-m,-n}^{(\mathcal{N}=2)}\rangle$ exist in the post-CDC theory (together forming a single $b_{3}$ eigenstate). The projection is then incorporated through the factor of $1/2$ in front of the first bracket.  In this connection, 
and as explained in the main body of the text, 
we emphasize that
implicit in the above 
are orbifold-twisted sectors that also have an ${\bf e}$ action.  However the radius-dependent shift in the Virasoro generators does not apply to these sectors either;  indeed, these 
sectors are supersymmetric and (along with the other twisted sectors) 
therefore make no net contribution to the partition function.  

Finally we note that since the Scherk-Schwarz phase induced for the worldsheet fermions on going around 
the $T_{2}$ compactification is $e^{2\pi iq_{e}}$, 
the spectrum of the $\mathcal{N}=2$ theory is independent of the sign 
$\pm\frac{1}{2}$ in $\mathbf{e}$. Therefore, just as in the real-fermion formalism, 
one may reverse the sign of the $\psi_{-}$ shift, giving
\begin{equation}
\mathbf{e'\cdot Q}~=~\frac{1}{2}\left(b_{+}^{\dagger}b_{+}-d_{+}^{\dagger}d_{+}+b_{-}^{\dagger}b_{-}-d_{-}^{\dagger}d_{-}\right)~=~ \left(Q_{13}+Q_{24}\right)~.
\end{equation}
In this formalism the vector components are represented as
\beq
V_{7} ~ \equiv ~ -\left(\frac{1}{4},\frac{1}{4}\right)~,~~~~~
b_{3} ~ \equiv ~ -\left(0,\frac{1}{2}\right)~,~~~~~
\mathbf{e'} ~ \equiv ~ \left(\frac{1}{2},\frac{1}{2}\right)~.
\eeq
Indeed, the minimal unit for consistency is a block containing two complex fermions that are overlapped by $V_7$ and ${\bf e}$ and only one which is overlapped by $b_3$.

Thus, to summarize the above procedure, 
we may evaluate each contribution in its relevant basis. 
This is ultimately because the $\mathbf{e}$-dependent part of the untwisted $\mathcal{N}=2$ sector is simply projected to one-half of its original value by the orbifold, while the twisted
sector is independent of $\mathbf{e}$.

\newpage
\bibliographystyle{unsrt}

\begin{thebibliography}{99}

\bibitem{Abel:2015oxa} 
  S.~Abel, K.~R.~Dienes and E.~Mavroudi,
  Phys.\ Rev.\ D {\bf 91}, no. 12, 126014 (2015)
  [arXiv:1502.03087 [hep-th]].


\bibitem{Abel:2016hgy}
  S.~Abel,
  JHEP {\bf 1611}, 085 (2016)
  [arXiv:1609.01311 [hep-th]].


\bibitem{Dienes:2002bg}
 K.~R.~Dienes, E.~Dudas and T.~Gherghetta,
  Phys.\ Rev.\ Lett.\  {\bf 91}, 061601 (2003)
  [hep-th/0210294].

\bibitem{Dienes:2004rt}
  K.~R.~Dienes, E.~Dudas and T.~Gherghetta,
  Pramana {\bf 62}, 219 (2004).

\bibitem{Burdman:2006tz}
  G.~Burdman, Z.~Chacko, H.~S.~Goh and R.~Harnik,
  JHEP {\bf 0702}, 009 (2007)
  [hep-ph/0609152].

\bibitem{Cohen:2015gaa}
  T.~Cohen, N.~Craig, H.~K.~Lou and D.~Pinner,
  JHEP {\bf 1603}, 196 (2016)
  [arXiv:1508.05396 [hep-ph]].

\bibitem{Angelantonj:2014dia}
  C.~Angelantonj, I.~Florakis and M.~Tsulaia,
  Phys.\ Lett.\ B {\bf 736}, 365 (2014)
  [arXiv:1407.8023 [hep-th]].

\bibitem{Hamada:2015ria}
  Y.~Hamada, H.~Kawai and K.~y.~Oda,
  Phys.\ Rev.\ D {\bf 92}, 045009 (2015)
  [arXiv:1501.04455 [hep-ph]].

\bibitem{Nibbelink:2015ena}
  S.~G.~Nibbelink,
  J.\ Phys.\ Conf.\ Ser.\  {\bf 631}, no. 1, 012077 (2015)
  [arXiv:1502.03604 [hep-th]].

\bibitem{Florakis:2015txa}
  I.~Florakis,
  J.\ Phys.\ Conf.\ Ser.\  {\bf 631}, no. 1, 012079 (2015)
  [arXiv:1502.07537 [hep-th]].

\bibitem{Ashfaque:2015vta}
  J.~M.~Ashfaque, P.~Athanasopoulos, A.~E.~Faraggi and H.~Sonmez,
  Eur.\ Phys.\ J.\ C {\bf 76}, no. 4, 208 (2016)
  [arXiv:1506.03114 [hep-th]].

\bibitem{Blaszczyk:2015zta}
  M.~Blaszczyk, S.~G.~Nibbelink, O.~Loukas and F.~Ruehle,
  JHEP {\bf 1510}, 166 (2015)
  [arXiv:1507.06147 [hep-th]].

\bibitem{Nibbelink:2015vha}
  S.~G.~Nibbelink, O.~Loukas and F.~Ruehle,
  Fortsch.\ Phys.\  {\bf 63}, 609 (2015)
  [arXiv:1507.07559 [hep-th]].

\bibitem{Angelantonj:2015nfa}
  C.~Angelantonj, I.~Florakis and M.~Tsulaia,
  Nucl.\ Phys.\ B {\bf 900}, 170 (2015)
  [arXiv:1509.00027 [hep-th]].

\bibitem{Satoh:2015nlc}
  Y.~Satoh, Y.~Sugawara and T.~Wada,
  JHEP {\bf 1602}, 184 (2016)
  [arXiv:1512.05155 [hep-th]].


\bibitem{Athanasopoulos:2016aws}
  P.~Athanasopoulos, A.~E.~Faraggi, S.~Groot Nibbelink and V.~M.~Mehta,
  JHEP {\bf 1604}, 038 (2016)
  [arXiv:1602.03082 [hep-th]].

\bibitem{Sugawara:2016lpa}
  Y.~Sugawara and T.~Wada,
  JHEP {\bf 1608}, 028 (2016)
  [arXiv:1605.07021 [hep-th]].

\bibitem{Kounnas:2016gmz}
  C.~Kounnas and H.~Partouche,
  Nucl.\ Phys.\ B {\bf 913}, 593 (2016)
  [arXiv:1607.01767 [hep-th]].

\bibitem{Florakis:2016ani}
  I.~Florakis and J.~Rizos,
  Nucl.\ Phys.\ B {\bf 913}, 495 (2016)
  [arXiv:1608.04582 [hep-th]].

\bibitem{Satoh:2016izo}
  Y.~Satoh and Y.~Sugawara,
  JHEP {\bf 1702}, 024 (2017)
  [arXiv:1611.08076 [hep-th]].


\bibitem{Kounnas:2017mad}
  C.~Kounnas and H.~Partouche,
  Nucl.\ Phys.\ B {\bf 919}, 41 (2017)
  [arXiv:1701.00545 [hep-th]].

\bibitem{Florakis:2017ecd}
  I.~Florakis and J.~Rizos,
  Nucl.\ Phys.\ B {\bf 921}, 1 (2017)
  [arXiv:1703.09272 [hep-th]].

\bibitem{Faraggi:2017cnh}
  A.~E.~Faraggi, J.~Rizos and H.~Sonmez,
  arXiv:1709.08229 [hep-th].

\bibitem{Coudarchet:2017pie}
  T.~Coudarchet, C.~Fleming and H.~Partouche,
  arXiv:1711.09122 [hep-th].

\bibitem{Mourad:2017rrl}
  J.~Mourad and A.~Sagnotti,
  arXiv:1711.11494 [hep-th].

\bibitem{BlumDienes}
  J.~D.~Blum and K.~R.~Dienes,
  Nucl.\ Phys.\ B {\bf 516}, 83 (1998)
  [hep-th/9707160].

\bibitem{DienesLennekSharma}
  K.~R.~Dienes, M.~Lennek and M.~Sharma,
  Phys.\ Rev.\ D {\bf 86}, 066007 (2012)
  [arXiv:1205.5752 [hep-th]].

\bibitem{Aaronson}
  B.~Aaronson, S.~Abel and E.~Mavroudi,
  Phys.\ Rev.\ D {\bf 95}, no. 10, 106001 (2017)
  [arXiv:1612.05742 [hep-th]].


\bibitem{missusy}
  K.~R.~Dienes,
  Nucl.\ Phys.\ B {\bf 429}, 533 (1994)
  [hep-th/9402006].

\bibitem{supertraces}
  K.~R.~Dienes, M.~Moshe and R.~C.~Myers,
  Phys.\ Rev.\ Lett.\  {\bf 74}, 4767 (1995)
  [hep-th/9503055].

\bibitem{Dienespath}
  K.~R.~Dienes,
  Phys.\ Rept.\  {\bf 287}, 447 (1997)
  [hep-th/9602045].

\bibitem{DDG1}
  K.~R.~Dienes, E.~Dudas and T.~Gherghetta,
  Phys.\ Lett.\ B {\bf 436}, 55 (1998)
  [hep-ph/9803466].

\bibitem{DDG2}
   K.~R.~Dienes, E.~Dudas and T.~Gherghetta,
  Nucl.\ Phys.\ B {\bf 537}, 47 (1999)
  [hep-ph/9806292].


\bibitem{DDG3}
  K.~R.~Dienes, E.~Dudas and T.~Gherghetta,
  hep-ph/9807522.


\bibitem{TV}
  T.~R.~Taylor and G.~Veneziano,
  Phys.\ Lett.\ B {\bf 212}, 147 (1988).


\bibitem{orbifoldguts}
  L.~J.~Hall and Y.~Nomura,
  Phys.\ Rev.\ D {\bf 64}, 055003 (2001)
  [hep-ph/0103125].

\bibitem{agashe}
  K.~Agashe,
  JHEP {\bf 0105}, 017 (2001)
  [hep-ph/0012182].

\bibitem{Kakushadze:1999bb} 
  Z.~Kakushadze and T.~R.~Taylor,
  Nucl.\ Phys.\ B {\bf 562}, 78 (1999)
  [hep-th/9905137].

\bibitem{Aaronson:2016kjm}
  B.~Aaronson, S.~Abel and E.~Mavroudi,
  Phys.\ Rev.\ D {\bf 95}, no. 10, 106001 (2017)
  [arXiv:1612.05742 [hep-th]].


\bibitem{FF1}
  H.~Kawai, D.~C.~Lewellen and S.~H.~H.~Tye,
  Nucl.\ Phys.\ B {\bf 288}, 1 (1987).


\bibitem{FF2}
  I.~Antoniadis, C.~P.~Bachas and C.~Kounnas,
  Nucl.\ Phys.\ B {\bf 289}, 87 (1987).


\bibitem{FF3}
  H.~Kawai, D.~C.~Lewellen, J.~A.~Schwartz and S.~H.~H.~Tye,
  Nucl.\ Phys.\ B {\bf 299}, 431 (1988).


 \bibitem{Ferrara:1987es}
  S.~Ferrara, C.~Kounnas and M.~Porrati,
  Nucl.\ Phys.\ B {\bf 304}, 500 (1988).


\bibitem{Ferrara:1987qp}
  S.~Ferrara, C.~Kounnas and M.~Porrati,
  Phys.\ Lett.\ B {\bf 206}, 25 (1988).


\bibitem{Ferrara:1988jx}
  S.~Ferrara, C.~Kounnas, M.~Porrati and F.~Zwirner,
  Nucl.\ Phys.\ B {\bf 318}, 75 (1989).


\bibitem{Kounnas:1989dk}
  C.~Kounnas and B.~Rostand,
  Nucl.\ Phys.\ B {\bf 341}, 641 (1990).


\bibitem{Antoniadis:1992fh}
  I.~Antoniadis, C.~Munoz and M.~Quiros,
  Nucl.\ Phys.\ B {\bf 397}, 515 (1993)
  [hep-ph/9211309].

\bibitem{Abel:2015rkm}
  S.~Abel and R.~J.~Stewart,
  JHEP {\bf 1602}, 182 (2016)
  [arXiv:1511.02880 [hep-th]].


\bibitem{Antoniadis:1990hb}
  I.~Antoniadis, G.~K.~Leontaris and J.~Rizos,
  Phys.\ Lett.\ B {\bf 245}, 161 (1990).


\bibitem{Faraggi:1991be}
  A.~E.~Faraggi,
  Phys.\ Lett.\ B {\bf 274}, 47 (1992).


\bibitem{Faraggi:1991jr}
  A.~E.~Faraggi,
  Phys.\ Lett.\ B {\bf 278}, 131 (1992).


\bibitem{Faraggi:1992fa}
  A.~E.~Faraggi,
  Nucl.\ Phys.\ B {\bf 387}, 239 (1992)
  [hep-th/9208024].


\bibitem{Faraggi:1994eu}
  A.~E.~Faraggi,
  Phys.\ Lett.\ B {\bf 339}, 223 (1994)
  [hep-ph/9408333].
  

\bibitem{Dienes:1995sv}
  K.~R.~Dienes and A.~E.~Faraggi,
  Phys.\ Rev.\ Lett.\  {\bf 75}, 2646 (1995)
  [hep-th/9505018].

\bibitem{Dienes:1995bx}
  K.~R.~Dienes and A.~E.~Faraggi,
  Nucl.\ Phys.\ B {\bf 457}, 409 (1995)
  [hep-th/9505046].


\bibitem{URL}
   http://www.maths.dur.ac.uk/$\sim$dma0saa/arXiv-plus/Dec2017/supp\_material


\bibitem{DSW}
  M.~Dine, N.~Seiberg and E.~Witten,
  Nucl.\ Phys.\ B {\bf 289}, 589 (1987).


\bibitem{upcoming}
  S.~A.~Abel and K.~R.~Dienes, to appear.

\bibitem{Chacko:2008cb}
  Z.~Chacko, C.~A.~Krenke and T.~Okui,
  JHEP {\bf 0901}, 050 (2009)
  [arXiv:0809.3820 [hep-ph]].

\bibitem{Kats:2017ojr}
  Y.~Kats, M.~McCullough, G.~Perez, Y.~Soreq and J.~Thaler,
  JHEP {\bf 1706}, 126 (2017)
  [arXiv:1704.03393 [hep-ph]].

\bibitem{Abel:2017rch}
  S.~Abel and R.~J.~Stewart,
  Phys.\ Rev.\ D {\bf 96}, no. 10, 106013 (2017)
  [arXiv:1701.06629 [hep-th]].


\bibitem{Jack:1989tv}
  I.~Jack and D.~R.~T.~Jones,
  Phys.\ Lett.\ B {\bf 234}, 321 (1990).


\bibitem{Garcia:2015sfa}
  I.~Garcia Garcia, K.~Howe and J.~March-Russell,
  JHEP {\bf 1512}, 005 (2015)
  [arXiv:1510.07045 [hep-ph]].


\end{thebibliography}

\end{document}